\documentclass[a4paper,fleqn]{cas-dc}
\usepackage{placeins}
\usepackage[sort,numbers]{natbib}
\usepackage{caption}
\usepackage{subcaption}
\usepackage{tikz}
\def\checkmark{\tikz\fill[scale=0.4](0,.35) -- (.25,0) -- (1,.7) -- (.25,.15) -- cycle;}
\usepackage{amsmath}
\usepackage{pifont}

\newcommand{\xmark}{\text{\ding{55}}}
\def\tsc#1{\csdef{#1}{\textsc{\lowercase{#1}}\xspace}}
\tsc{WGM}
\tsc{QE}
\tsc{EP}
\tsc{PMS}
\tsc{BEC}
\tsc{DE}

\begin{document}
\let\WriteBookmarks\relax
\def\floatpagepagefraction{1}
\def\textpagefraction{.001}

\shorttitle{Deep Learning for Cognitive Radio Networks}


\title [mode = title]{Deep Learning Frameworks for Cognitive Radio Networks: Review and Open Research Challenges} 




%
\author[1]{Senthil Kumar Jagatheesaperumal}[type=editor,
                        auid=000,bioid=1,
                        orcid=0000-0002-9516-0327]



\ead{senthilkumarj@mepcoeng.ac.in}


\cormark[1]
\address[1]{Department of Electronics and Communication Engineering, Mepco Schlenk Engineering College, Sivakasi, Tamil Nadu, India}

\author[2]{Ijaz Ahmad}[style=chinese]
\ead{ijaz.ahmad@vtt.fi}

\address[2]{VTT Technical Research Centre of Finland, Finland}

\author[3]{Marko Höyhtyä}[%
   ]
\ead{firstname.lastname@vtt.fi}

\address[3]{VTT Technical Research Centre of Finland, Finland}


\author[4]{Suleman Khan}[%
   ]
\ead{suleman.khan@liu.se}
\address[4]{Linköping University,Sweden}

\author[5]{Andrei Gurtov}[%
   ]
\ead{gurtov@acm.org}

\address[5]{Linköping University,Sweden}
	
\cortext[cor1]{Corresponding author:Ijaz Ahmad, VTT Technical Research Centre of Finland, Finland, Email ID:ijaz.ahmad@vtt.fi}



\begin{abstract}
Deep learning has been proven to be a powerful tool for addressing the most significant issues in cognitive radio networks, such as spectrum sensing, spectrum sharing, resource allocation, and security attacks. The utilization of deep learning techniques in cognitive radio networks can significantly enhance the network's capability to adapt to changing environments and improve the overall system's efficiency and reliability. As the demand for higher data rates and connectivity increases, B5G/6G wireless networks are expected to enable new services and applications significantly. Therefore, the significance of deep learning in addressing cognitive radio network challenges cannot be overstated. This review article provides valuable insights into potential solutions that can serve as a foundation for the development of future B5G/6G services. By leveraging the power of deep learning, cognitive radio networks can pave the way for the next generation of wireless networks capable of meeting the ever-increasing demands for higher data rates, improved reliability, and security.


\end{abstract}




\begin{keywords}
Cognitive radio network \sep Deep learning \sep Machine learning \sep Spectrum awareness \sep Resource allocation \sep Security.
\end{keywords}

\maketitle

\section{Introduction}
\label{introduction}


Cognitive Radio Networks (CRN) based wireless communication systems are programmed to select the most suitable channel among available options to optimize the use of spectrum resources. \textcolor{black}{The cognitive radio (CR), founded on the principles of software radio, significantly improves personal communication by employing model-based reasoning alongside a knowledge representation language\cite{haykin2005cognitive}. This approach allows for the intelligent adaptation of radio protocols and spectrum utilization, effectively aligning with user requirements in real-time. Such systems are characterized by their ability to adapt to environmental conditions, ensuring reliable communication and optimizing spectrum usage~\cite{mitola1999cognitive}. It emphasizes critical components such as radio-scene analysis, channel-state estimation, and dynamic spectrum management. } Typical CRN applications are opportunistic spectrum utilization, intelligent beam-forming, automated interoperability, communication among emergency services, spectrum trading, and so on. CRN's rise in popularity in industries and academia makes them one of the key enablers for modern-day networks. Currently, the radio spectrum is allocated to different wireless technologies. While the frequency spectrum in certain frequency bands of mobile communications networks is getting crowded, there are frequencies that are hardly used, which results in spectral inefficiency~\cite{serrano2021random}. The increasing number of mobile devices and the limited availability of the radio spectrum are making it difficult for existing wireless technologies to survive in current market trends. CRN enables solving the challenges existing in spectrum scarcity by enabling opportunistic use of the radio spectrum. With the support of CR, even unlicensed users can utilize the licensed frequency bands from inactive users and currently not utilizing their allocated frequency spectrum. The use of Software Defined Radios (SDRs) allows CR systems to adapt dynamically to the changes in their operating environments~\cite{thameur2021sdr}. These SDRs can sense the under and over-utilized parts of the spectrum and redistribute the transmission to ensure that the spectrum is more efficiently utilized. Systems thrive on an understanding by building methodology. This, in turn, increases spectral efficiency, operational flexibility, and adaptability to the environments and improves the interoperability between existing wireless networks.

%
CRN is subjected to more risks and challenges in terms of security threats (For example, what threat or attacks and their consequences), the raising of false alarms by secondary users, and the detection of idle resources~\cite{salahdine2020security}. Better sensing-aware routing protocols are needed in CRN networks to improve throughput in interference-limited environments. Interference management of CRN users using an unlicensed spectrum with the users in the licensed spectrum needs to be carefully dealt with, especially when they are involved in emergency services~\cite{salameh2020intelligent}. Security risks also arise from using licensed user emulation and illegally downloading malicious software. 
A group of researchers set out on a mission to solve the challenging problems that plague the security of the network in the realm of CRN. They explored a few of the popular concerns such as the complexities of primary user emulation attacks~\cite{batool2022detection}, spectrum sensing data falsification~\cite{kumar2022hierarchical}, and the constant risk of jamming attacks~\cite{thien2021transfer}. They create a ground-breaking approach that incorporates state-of-the-art techniques to precisely detect and stop these criminal activities through ceaseless experimentation and inventive algorithms. The outcomes from their research open the door for cutting-edge methods of spectrum authentication and encryption that guarantee the integrity and equitable distribution of resources in CRN.


It is important to note that deep learning (DL) is a relatively new paradigm for its implication in CRN. Due to its capacity to automatically learn complex patterns and characteristics from huge datasets, DL-based solutions are preferred for addressing security concerns in CRN. This enables the accurate detection of abnormalities and threats. DL models' depth and complexity provide robust and adaptable defenses, making them well-suited to deal with dynamic threats in cognitive radio environments. This has made it possible to accurately detect and categorize signals, assaults, and abnormalities. In order to effectively meet the issues in the CRN environment, it has the capacity to learn hierarchical representations and adaptively update models. DL-inspired systems and their associated technology currently span a huge set of significant applications for driving CRN. Such applications include efficient resource allocation~\cite{liu2018deep}, Spectrum occupancy reconstruction~\cite{hlophe2019spectrum}, Multi-efficiency resource allocation~\cite{liu2018multi}, Device fingerprinting~\cite{merchant2018deep}, Signal  classification~\cite{liu2020data}, Modulation classification~\cite{kim2021deep,tang2018digital} and Modulation recognition~\cite{wang2019data}. CRN solutions addressing the challenges faced in the Internet of Things (IoT) are reviewed in~\cite{rawat2016cognitive} with a focus on energy, scalability, equipment capability, and environments.

One of the primary challenges in CRN is to meet the demand for radio resources—most of the sub-networks in CRN attempt to manage their requirements effectively. But effective coordination of resources requires specialized frameworks. Moreover, exclusive frameworks enhance spectral characteristics, communication capability, and improved security. Particularly data-driven DL frameworks are used for numerous applications of CRN. A few of the most promising applications are elaborated on with appropriate examples in the rest of this section. 

\subsection{Our Contributions}
This survey aims to provide a thorough review of the usage of DL frameworks for CRN application areas. With this scope, we attempt to answer the following key issues in the CRN domain:

\begin{itemize}
	\item Why is DL a promising approach to solving the challenges in CRN?
	\item We summarize recent and promising outcomes of the CRN evolved through the incorporation of DL techniques.
	\item Finally, we present the future research directions for using DL frameworks in CRN research. 
\end{itemize}

\begin{figure*}
  \centering \includegraphics[width=\textwidth]{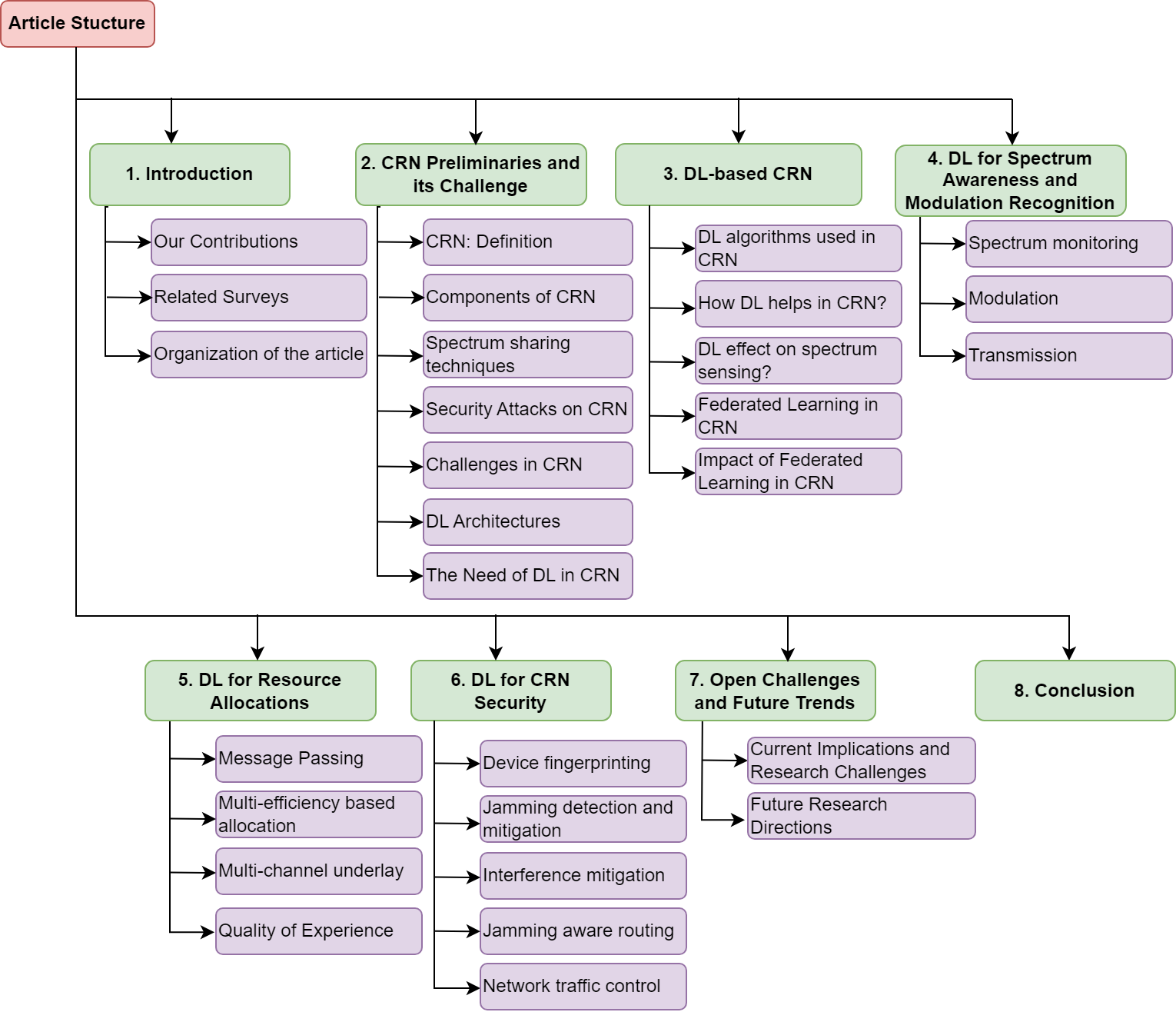}
   \caption{\textcolor{black}{Overall organization of the article.}}
   \label{fig:struct}
\end{figure*}

The existing reviews, articles, and other magazines only partially address the above-mentioned key issues. This survey article focuses on previous works and addresses the overlap between DL and CRN application domains. \textcolor{black}{The structure of the article is presented in Figure~\ref{fig:struct}.}

\subsection{Related Surveys} 
\begin{figure}
  \centering \includegraphics[width=0.50\textwidth]{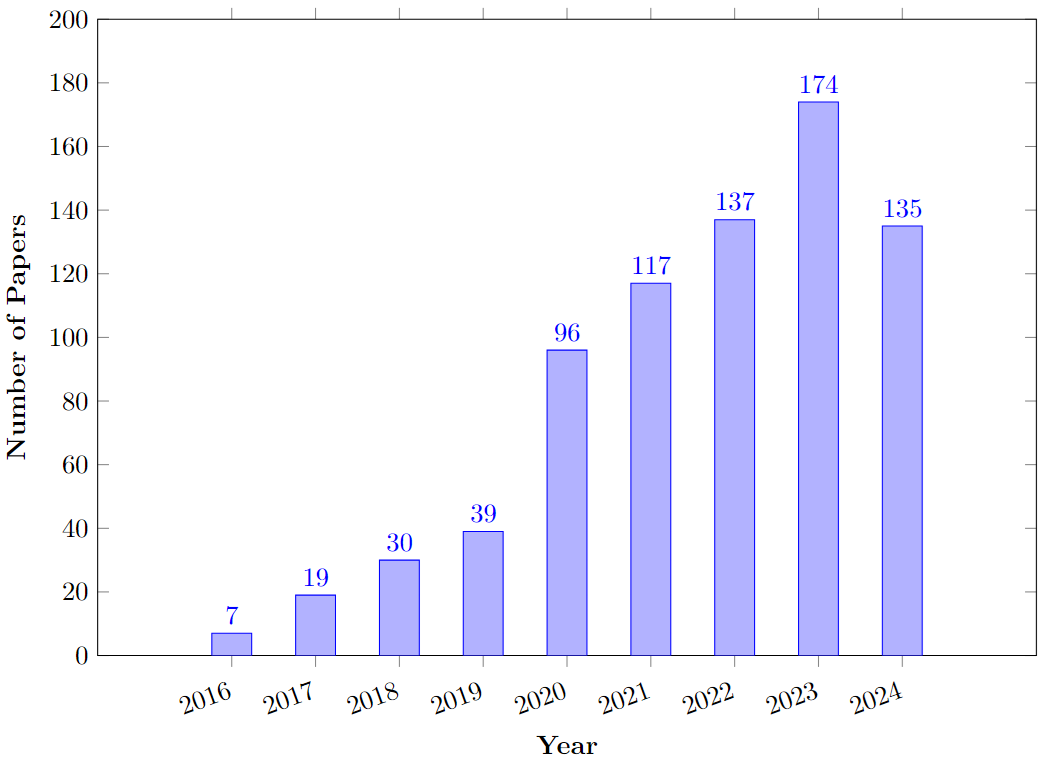}
   \caption{\textcolor{black}{Number of related articles published per year (2016-2024 (September))}}
   \label{fig:paper}
\end{figure}

In recent research, many solutions have been proposed to overcome the challenges faced by CRN in different application scenarios. Few survey articles have reviewed those proposed solutions under different categories of scopes, as shown in Table~\ref{tab:survey}. Progress of research up to date is summarized from the review articles on CRN. From the time-frequency-space perspective, Ding~\textit{et al.}~\cite{ding_spectrum_2017} performed a comparative analysis of various spectrum interference algorithms from literature and summarized them with key research challenges. Gupta \textit{et al.}~\cite{gupta2019progression} presented a survey with a focus on the state-of-the-art development of spectrum sensing and its taxonomy. \textcolor{black}{Figure~\ref{fig:paper} shows the bar graph of related published articles from the year 2016-2024 (September)}

Moreover, this work investigated the modeling aspects of different spectrum sensing schemes. Likely, in~\cite{muzaffar2024review}, a review of numerical analysis on spectrum sensing techniques is performed. Moreover, it described the key role of matched filters, energy, and detectors in the performance analysis of the spectrum sensing methods for CRN. Further, Khasawneh~\textit{et al.}~\cite{khasawneh2023convergence} presented the compilation of QoS provisioning for spectrum sensing. As such, here for spectrum sensing, QoS approaches were also investigated along with decision components, spectrum sharing, and mobility components. The work in~\cite{bouabdellah2018network} surveyed related literature in terms of classification of attacks targeting the network layer, and while Salahdine~\textit{et al.}~\cite{salahdine2020security} provided a systematic review on the classification and analysis of physical layer attacks. Both the former and later works focused on countermeasures of attacks on physical and network layers, respectively.

Few papers have reviewed recent works that map and integrate machine learning (ML) and DL into the CRN domain. A more recent survey by Shekhawat~\textit{et al.}~\cite{shekhawat2021review} reviewed the deep spectrum sensing for 5G networks, which incorporated DL frameworks. The article discussed the standardization of deep sensing techniques for the next generation of massive connectivity and addressed the issues related to cooperation overhead. Wang~\textit{et al.}~\cite{wang2019survey} conducted an RL-based spectrum allocation survey. They analyzed the dynamic spectrum allocation along with its merits, demerits, and potential applications. In another similar work, Di~\textit{et al.}\cite{felice2019reinforcement} summarized the research findings on RL-based spectrum management analyzed over cooperative as well as mixed cooperative tasks and Babu~\textit{et al.}\cite{babu2019survey} reviewed on optimization of resources and spectrum allocation using AI techniques. This AI-enabled feature helps to identify favorable conditions for applying the techniques. In a recent work by Bharathy~\textit{et al.}~\cite{bharathy2021research}, the authors summarized the cost-effective CRN services and research \& development in the CRN domain. Besides, the numerical design principles involved in analyzing the spectrum utilization were also investigated, considering the features of spectrum sensing. Very recently, Kaur et al. \cite{kaur2022comprehensive} summarized the impact of machine learning techniques on dynamic and intelligent spectrum management. From the intelligence point of view, deep learning  techniques could play an important role in investigating CRN in a diversified range of applications.

\begin{figure*}
  \centering \includegraphics[width=1\textwidth]{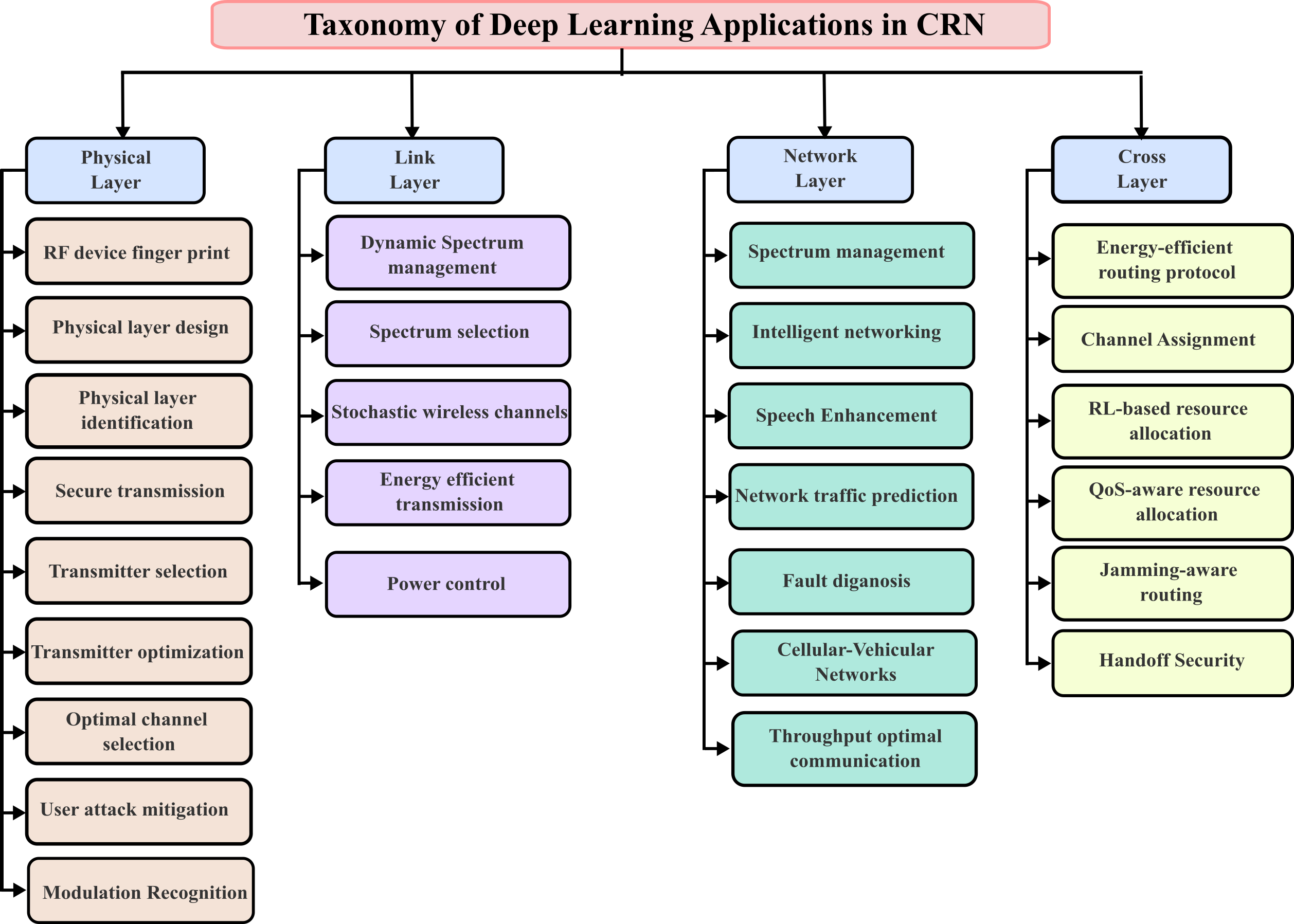}
   \caption{Taxonomy of DL Applications in CRN.} 
   \label{fig:taxanomy}
\end{figure*}

In summary, it can be concluded that some existing surveys compiled the research in the CRN only from the single perspective of managing the resources and securing the systems. Some surveys in the past few years have introduced the implications of DL from a high level and failed to address their impact on CRN comprehensively. With the growing enthusiasm from the research community in recent years, a large number of new research results have emerged with the deployment of DL in CRN. Although the existing surveys listed in Table~\ref{tab:survey} have received the CRN from various perspectives, none of them focuses on the issues in CRN from the DL perspective in a comprehensive way. This motivates us to present a systematic review from multiple perspectives, including resource allocation, DL architectures, security, research issues, and applications, to provide researchers with an informative and updated viewpoint. Fig.~\ref{fig:taxanomy} summarizes the taxonomy of DL applications in CRN based on the various layers of operations.

{\color{black}
\begin{table*}
\scriptsize
\caption{Summary of existing surveys in Cognitive Radio Network} 
\label{tab:survey}
\begin{tabular}{|>{\arraybackslash}p{2cm}|>{\centering\arraybackslash}p{0.5cm}|>{\centering\arraybackslash}p{0.8cm}|>{\centering\arraybackslash}p{0.8cm}|>{\centering\arraybackslash}p{0.8cm}|>{\centering\arraybackslash}p{0.8cm}|>{
\centering\arraybackslash}p{0.8cm}|>{\centering\arraybackslash}p{0.7cm}|>{\centering\arraybackslash}p{0.7cm}|>{\centering\arraybackslash}p{0.8 cm}|>{\arraybackslash}p{5cm}|}
 \hline
\textbf{Refs. (Author)}
& \textbf{Year}
& \textbf{\tiny Spectrum Sensing} 
& \textbf{\tiny Spectrum Sharing} 
& \textbf{\tiny Spectrum Handoff} 
& \textbf{\tiny Security \& Privacy}
& \textbf{\tiny PU Classification}
& \textbf{\tiny Routing \newline Protocols}
& \textbf{\tiny Energy\& Solution}
& \textbf{\tiny DL Approaches}
& \textbf{\tiny Key contribution}
 \\ \hline \hline
Thomas\textit{et al.}\cite{thomas2017survey} & 2017 & \xmark{} & \xmark{} & \checkmark{} & \xmark{} & \xmark{} & \xmark{} & \xmark{} & \xmark{} & Handoff schemes and modeling techniques of PU.  \\ \hline
Akram\textit{et al.}\cite{akram2017review} & 2017 & \xmark{} & \xmark{} & \xmark{} & \checkmark{} & \xmark{} & \xmark{} & \xmark{} & \xmark{} &  Explanation on the consequences of attacks.  \\ \hline
Bouabdellah\textit{et al.}\cite{bouabdellah2018network} & 2018 & \xmark{} & \xmark{} & \xmark{} & \checkmark{} & \checkmark{} & \xmark{} & \xmark{} & \xmark{} &  Attacks targeting the network layer functionalities are classified.  \\ \hline
Mir\textit{et al.}\cite{mir2018time} & 2018 & \xmark{} & \xmark{} & \checkmark{} & \xmark{} & \xmark{} & \xmark{} & \xmark{} & \xmark{} & Classification of time triggered handoff schemes.   \\ \hline
Osman\textit{et al.}\cite{osman2018survey} & 2018 & \xmark{} & \xmark{} & \xmark{} & \xmark{} & \xmark{} & \xmark{} & \xmark{} & \xmark{} & Clustering algorithms for Adhoc networks.   \\ \hline
Rahamathullah\textit{et al.}\cite{rahamathullah2018review} & 2018 & \xmark{} & \checkmark{} & \xmark{} & \xmark{} & \xmark{} & \xmark{} & \xmark{} & \xmark{} &  Fuzzy and ML-based approaches. \\ \hline
Yin\textit{et al.}\cite{yin2018spectrum} & 2018 & \xmark{} & \xmark{} & \xmark{} & \xmark{} & \xmark{} & \xmark{} & \xmark{} & \xmark{} & Spectrum Utilization in industrial WSN.  \\ \hline
Hu\textit{et al.}\cite{hu2018full} & 2018 & \checkmark{} & \checkmark{} & \xmark{} & \xmark{} & \xmark{} & \xmark{} & \xmark{} & \xmark{} &  Full spectrum sharing in 5G scenario.  \\ \hline
Amjad\textit{et al.}\cite{amjad2018wireless} & 2018 & \xmark{} & \xmark{} & \xmark{} & \checkmark{} & \xmark{} & \xmark{} & \xmark{} & \xmark{} & Wireless multimedia QoE and security are analyzed.  \\ \hline
Kakalou\textit{et al.}\cite{kakalou2018survey} & 2018 & \checkmark{} & \xmark{} & \xmark{} & \xmark{} & \xmark{} & \xmark{} & \xmark{} & \xmark{} & Compressive sensing and sub-Nyquist techniques are analysed. \\ \hline
Sharmila\textit{et al.}\cite{sharmila2019spectrum} & 2019 & \xmark{} & \checkmark{} & \xmark{} & \xmark{} & \xmark{} & \xmark{} & \xmark{} & \xmark{} & Manifold techniques and hybrid spectrum access technology.  \\ \hline
Gupta\textit{et al.}\cite{gupta2019progression} & 2019 & \checkmark{} & \xmark{} & \xmark{} & \xmark{} & \xmark{} & \xmark{} & \xmark{} & \xmark{} & State-of-the-art and taxonomy of spectrum sensing.   \\ \hline
Wang\textit{et al.}\cite{wang2019survey} & 2019 & \xmark{} & \checkmark{} & \xmark{} & \xmark{} & \xmark{} & \xmark{} & \xmark{} & \xmark{} & RL-based spectrum allocation approaches. \\ \hline
Babu\textit{et al.}\cite{babu2019survey} & 2019 & \xmark{} & \xmark{} & \xmark{} & \xmark{} & \xmark{} & \xmark{} & \xmark{} & \checkmark{} &  Resource optimization and spectrum allocation using AI techniques.   \\ \hline
Ntshabele\textit{et al.}\cite{ntshabele2019energy} & 2019 & \xmark{} & \xmark{} & \xmark{} & \xmark{} & \xmark{} & \xmark{} & \checkmark{} & \xmark{} &  Clustering topologies and energy consumption.  \\ \hline
Benazzouza\textit{et al.}\cite{benazzouza2019survey} & 2019 & \checkmark{} & \xmark{} & \xmark{} & \xmark{} & \xmark{} & \xmark{} & \xmark{} & \xmark{} & Compressive spectrum sensing for wideband CRN.   \\ \hline
Elrhareg\textit{et al.}\cite{elrhareg2019routing} & 2019 & \xmark{} & \xmark{} & \xmark{} & \xmark{} & \checkmark{} & \xmark{} & \xmark{} & \xmark{} &  Domain applications of routing protocols.  \\ \hline
Tarek\textit{et al.}\cite{tarek2020survey} & 2020 & \xmark{} & \checkmark{} & \xmark{} & \xmark{} & \xmark{} & \xmark{} & \xmark{} & \xmark{} &   Packets scheduling and channel allocation for IoT applications.\\ \hline
Salahdine\textit{et al.}\cite{salahdine2020security} & 2020 & \xmark{} & \xmark{} & \xmark{} & \checkmark{} & \checkmark{} & \xmark{} & \xmark{} & \xmark{} &  Types of physical layer attacks are described and analyzed.  \\ \hline
Rao\textit{et al.}\cite{rao2020full} & 2020 & \xmark{} & \xmark{} & \xmark{} & \xmark{} & \xmark{} & \xmark{} & \checkmark{} & \xmark{} & Energy harvesting in full-duplex communication.   \\ \hline
Raj\textit{et al.}\cite{raj2020survey} & 2020 & \xmark{} & \xmark{} & \xmark{} & \xmark{} & \checkmark{} & \xmark{} & \xmark{} & \xmark{} & RL-based spectrum aware routing.   \\ \hline
Dhawan\textit{et al.}\cite{dhawan2020routing} & 2020 & \xmark{} & \xmark{} & \xmark{} & \checkmark{} & \xmark{} & \checkmark{} & \xmark{} & \xmark{} &  Vulnerability on routing protocols.  \\ \hline
Salih\textit{et al.}\cite{salih2020smart} & 2020 & \xmark{} & \xmark{} & \xmark{} & \xmark{} & \checkmark{} & \xmark{} & \xmark{} & \xmark{} &  Taxonomy of routing protocols for mobile CRN.  \\ \hline
Naikwadi\textit{et al.}\cite{naikwadi2020survey} & 2020 & \xmark{} & \xmark{} & \xmark{} & \xmark{} & \xmark{} & \xmark{} & \xmark{} & \checkmark{} & ANN based Spectrum inference and occupancy prediction.    \\ \hline
Diab\textit{et al.}\cite{diab2020survey} & 2020 & \xmark{} & \xmark{} & \xmark{} & \xmark{} & \xmark{} & \checkmark{} & \checkmark{} & \xmark{} &  IoT and Energy-Constrained frameworks  \\ \hline
Shekhawat\textit{et al.}\cite{shekhawat2021review} & 2021 & \checkmark{} & \xmark{} & \xmark{} & \xmark{} & \xmark{} & \xmark{} & \xmark{} & \checkmark{} &  Deep Spectrum Sensing for 5G networks \\ \hline
Jiang\textit{et al.}\cite{jiang2021location} & 2021 & \xmark{} & \xmark{} & \xmark{} & \checkmark{} & \xmark{} & \xmark{} & \xmark{} & \xmark{} &  Location privacy risks and threats in CRN components. \\ \hline
Kaur\textit{et al.}\cite{kaur2022comprehensive} & 2022 & \xmark{} & \xmark{} & \xmark{} & \xmark{} & \xmark{} & \xmark{} & \xmark{} & \xmark{} & Network coding schemes taxonomy.  \\ \hline
Bhattacharjee\textit{et al.}\cite{bhattacharjee2022cognitive} & 2022 & \xmark{} & \xmark{} & \xmark{} & \xmark{} & \xmark{} & \xmark{} & \xmark{} & \xmark{} & Spectrum access exclusive-use trading approaches.  \\ \hline
Manco\textit{et al.}\cite{manco2022spectrum} & 2022 & \checkmark{} & \checkmark{} & \xmark{} & \xmark{} & \xmark{} & \xmark{} & \checkmark{} & \checkmark{} &  Spectrum Sensing using software-defined radio. \\ \hline
Ul\textit{et al.}\cite{ul2022differential} & 2022 & \checkmark{} & \checkmark{} & \xmark{} & \checkmark{} & \xmark{} & \xmark{} & \checkmark{} & \checkmark{} &    Differential privacy issues in CRN.\\ \hline
\textcolor{black}{{Tolley\textit{et al.}\cite{tolley2023systematization}}} & \textcolor{black}{2023} & \checkmark{} & \checkmark{} & \checkmark{} & \xmark{} & \xmark{} & \xmark{} & \xmark{} & \xmark{} & \textcolor{black}{Spectrum sharing for satellite communications.}  \\ \hline
\textcolor{black}{Mohammadi\textit{et al.}\cite{mohammadi2023comprehensive}} & \textcolor{black}{2023} & \xmark{} & \xmark{} & \xmark{} & \xmark{} & \xmark{} & \xmark{} & \xmark{} & \xmark{} & \textcolor{black}{Full-duplex communication network architectures.}  \\ \hline
\textcolor{black}{Khasawneh\textit{et al.}\cite{khasawneh2023convergence}} & \textcolor{black}{2023} & \checkmark{} & \xmark{} & \xmark{} & \xmark{} & \xmark{} & \xmark{} & \xmark{} & \xmark{} & \textcolor{black}{QoS provisioning for spectrum sensing}  \\ \hline
\textcolor{black}{Sandya\textit{et al.}\cite{sandya2024real}} & \textcolor{black}{2024} & \xmark{} & \xmark{} & \xmark{} & \xmark{} & \checkmark{} & \xmark{} & \xmark{} & \xmark{} & \textcolor{black}{PU detection and TV white space Information acquisition }   \\ \hline
\textcolor{black}{Sharma\textit{et al.}\cite{sharma2024multi}} & \textcolor{black}{2024} & \xmark{} & \checkmark{} & \xmark{} & \xmark{} & \xmark{} & \xmark{} & \xmark{} & \xmark{} &  \textcolor{black}{Multi-objective optimization for spectrum sharing.}  \\ \hline
\textcolor{black}{Muzaffar\textit{et al.}\cite{muzaffar2024review}} & \textcolor{black}{2024} & \checkmark{} & \xmark{} & \xmark{} & \xmark{} & \xmark{} & \xmark{} & \xmark{} & \xmark{} & \textcolor{black}{Numerical analysis on spectrum sensing techniques.}   \\ \hline
Our Survey & \textcolor{black}{2024} & \checkmark{} & \checkmark{} & \checkmark{} & \checkmark{} & \checkmark{} & \checkmark{} & \checkmark{} & \checkmark{} & DL approaches for CRN considering sensing, sharing, PU classification, routing, security, and energy issues.   \\ \hline
\end{tabular}
\end{table*}
}
\subsection{Organization of the article}
This article is organized as follows: Section~\ref{sec:preliminaries} provides an overview of related work on CRN and on the characteristics of CRN and its associated challenges. Section~\ref{sec:DLforCRN} focuses on the DL frameworks with the summaries of recent survey articles contributed towards the CRN research domain and elaborates on the preliminary background of DL networks. Section~\ref{sec:spectrum} provides deep insight into the usage of DL frameworks for spectrum monitoring applications, modulation classification, and cognitive transmission. Section~\ref{sec:resource} presents the deployment strategies of DL frameworks for resource allocation tasks in CRN. The section~\ref{sec:security} offers the role of DL in addressing the network security issues of CRN. Section~\ref{sec:issues} elaborates on the open issues, challenges, and future scope of research in the prescribed domain and summarizes the lessons learned in this work with the key takeaway for researchers. Section~\ref{sec:conclusion} concludes the article. 
Table~\ref{table:terms} presents a list of acronyms used in this paper.
\begin{table}
\centering
\caption{A summary of acronyms used in this article} \vspace{9pt}
\label{table:terms}
\begin{tabular}{|p{2cm}|p{5cm}|}
\hline
Acronym & Description                                            \\ \hline \hline
ACO             & Ant Colony Optimisation                        \\ \hline
ACGAN           & Auxiliary Classifier Generative Adversarial Networks \\ \hline
ANN             & Artificial Neural Network                      \\ \hline
CR              & Cognitive Radio                                \\ \hline
CRN             & Cognitive Radio Networks                       \\ \hline
CNN             & Convolutional Neural Network                   \\ \hline
CSI             & Channel State Information                      \\ \hline
DDQN            & Double Deep Q-Network                          \\ \hline    
DL              & Deep Learning                                  \\ \hline
DNN             & Deep Neural Network                            \\ \hline
DoS             & Denial of Service                              \\ \hline
DRL             & Deep Reinforcement Learning                    \\ \hline
DQN             & Deep Q-Network                                 \\ \hline    
DSN             & Deep Stacking Network                          \\ \hline
GA              & Genetic Algorithm                              \\ \hline
GAN             & Generative Adversarial Networks                \\ \hline
GRN             & Gated Recurrent Network                        \\ \hline
IP              & Internet Protocol                              \\ \hline
IoT             & Internet of Things                             \\ \hline
NOMA            & Non-Orthogonal Multiple Access                 \\ \hline
LSTM            & Long Short-Term Memory                         \\ \hline
OFDM            & Orthogonal frequency-division multiplexing     \\ \hline
PCA             & Principal Component Algorithm                  \\ \hline
PSO             & Particle Swarm Optimization                    \\ \hline
PU              & Primary User                                   \\ \hline
QoS             & Quality of Service                             \\ \hline
QoE             & Quality of Experience                          \\ \hline
QAM             & Quadrature Amplitude Modulation                \\ \hline
ReLU            & Rectified Linear Unit                          \\ \hline
RNN             & Recurrent Neural Network                       \\ \hline
SDR             & Software Defined Radio                         \\ \hline
SVM             & Support Vector Machine                         \\ \hline
VC              & Varying Channel                                \\ \hline
WiFi            & Wireless Fidelity                              \\ \hline
\end{tabular}
\end{table}


\section{CRN Preliminaries and its Challenges} 
\label{sec:preliminaries}


This section provides the basics of CRN and its role in modern wireless communication networks. It also provides an overview of the core components and characteristics of CRN. 

\subsection{CRN: Definition}
CRNs are networks of smart radios that possess the ability to sense the external environment, make intelligent decisions by learning from history and adjust their transmission parameters accordingly, based on the environmental states. They dynamically adapt to the usage of the available spectrum by detecting and utilizing underutilized frequency bands. This allows for efficient utilization of the scarce radio frequency spectrum and enables better utilization of the available bandwidth for wireless communication ~\cite{4481339, 7266528}. The licensed spectrum band of frequencies is utilized exclusively by designated users. CRN utilizes the unlicensed band of frequency and enables the users to follow certain rules based on time, location, transmission power, and accessing the limit of the frequency bands~\cite{obite2021overview}. Spectrum sensing mechanisms effectively observe and determine the available portion of the spectrum. The spectrum decision process selects the best available channel in the identified available spectra. Coordinated access to the selected channel with other authenticated users is carried out through effective spectrum-sharing mechanisms. On recognition of licensed users (also called primary users), the secondary users may vacate the channel and move to the next best channel in the spectrum through the process of spectrum mobility. 


Secondary users carry out this dynamic spectrum access by deftly accessing the spectrum holes throughout the entire spectrum with improved performance without detrimental interference to the primary users~\cite{el2021performance}. CRN based on smart wireless networks, based on the interactions with the surroundings, is capable of altering its operational parameters for establishing intelligent jamming-aware routing, spectrum management, optimizing the spectrum sensing strategies, and effective management of sensing-throughput trade-offs.

\begin{figure*}
  \centering \includegraphics[width=0.75\textwidth]{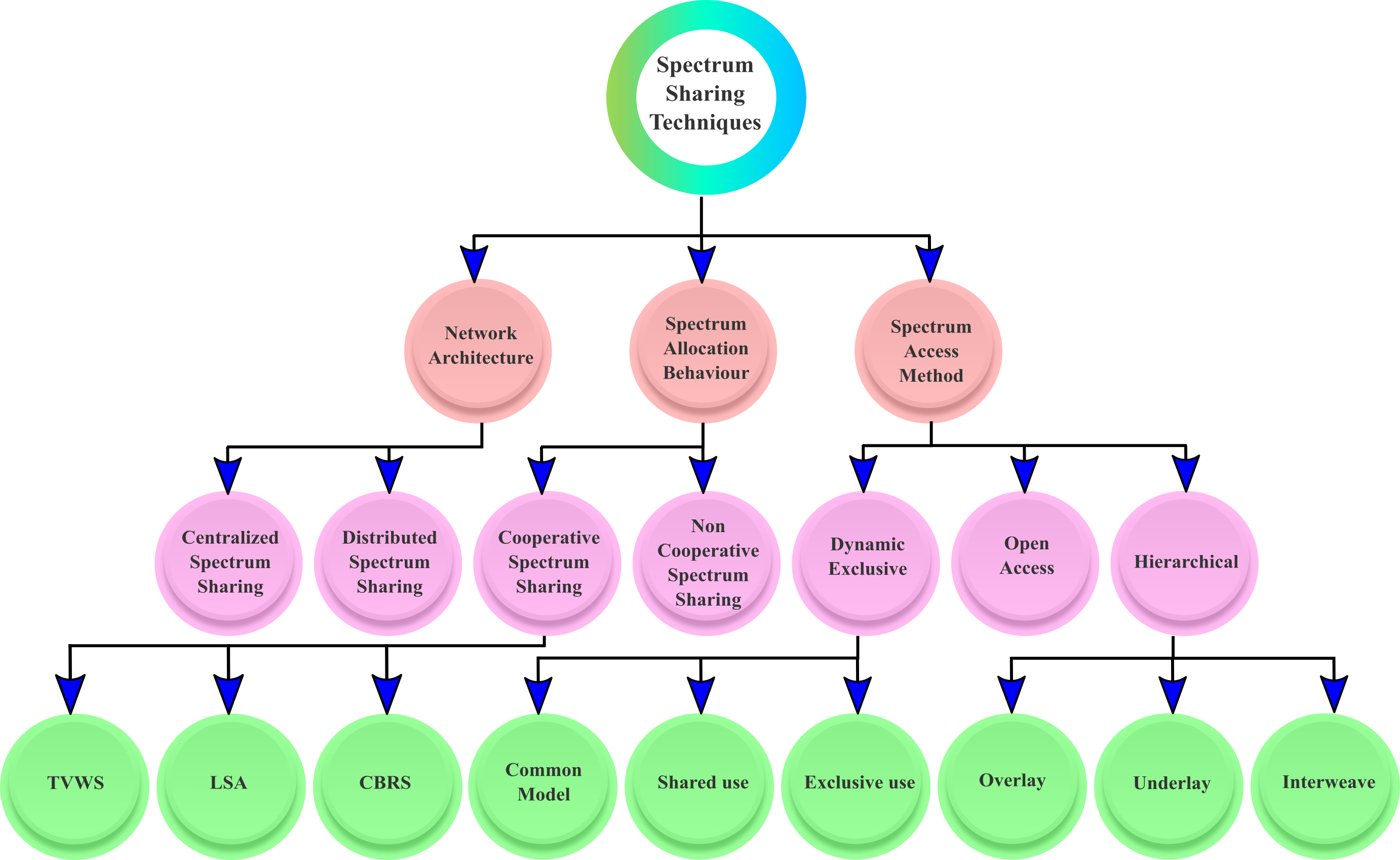}
   \label{fig:spreadtechniques}
\end{figure*}

For better utilization of the inherent features of the CRN smart wireless systems, they possess three main characteristics based on their capabilities for reconfigurability and self-organizing features of the network.

\begin{itemize}
    \item \textit{Cognitive capability:} In the features pertained to the cognitive capability of the network, they mainly address the spectrum sensing strategies, location identification, and discovery of networks and services.
    
    \item \textit{Re-configurable capability:} The re-configurable features address the issues related to frequency agility, dynamic frequency selection, Adaptive modulation coding, Transmission power control, and access mechanisms followed in the network of dynamic CRN systems.
    
    \item \textit{Self-organizing capability:} With intelligence loaded to the communication terminal devices employed for CRN, they should be able to organize their communication strategies themselves based on their sensing of the dynamic environment.
    
\end{itemize}

\subsection{Components of CRN}

\subsubsection{Cognitive radios} 
Cognitive radio is essentially a software-defined radio or an intelligent radio system that can be programmed to dynamically and at run-time adjust to specific frequency bands. As mobile frequency bands are getting crowded, with cognitive radio, we can utilize any of the available frequency bands for communication. Its architecture includes hardware, software, and cognitive modules, bound to a set of design rules and functions. Cognitive radio takes responsibility for addressing a large range of challenges in CRN such as spectrum sharing problems, power control, modulation recognition, etc~\cite{kulin2018end,wang2019data}. They are also involved in provisioning adaptive spectrum sharing, and multi-band spectrum sensing with the support of evolutionary algorithms~\cite{tuberquia2018new}.


SDR is a radio communication system that is responsible for addressing physical layer issues. They act as one of the core components in driving digital wireless technology. 
SDRs typically consist of hardware components, including amplifiers, filters, detectors, modulators, and demodulators, as well as software-based functionalities such as modulation, signal processing, channel coding, and data encapsulation.  These software-based functionalities are typically implemented on a personal computer or embedded systems. They assist in listening to broadcast stations and receiving satellite signals, aircraft radio signals, and other medium-wave radio signals. The basic architecture of the SDR includes antennas, low noise amplifiers, filters, mixers, analog to digital converters, and in-phase (I) and quadrature (Q) samples. The computer running the SDR software, will read the IQ components and decode them and let the users acquire the data stream. In~\cite{stamou2019autonomic}, the authors proposed an Autonomic Network Management framework through network virtualization demonstrated for SDR-based CRN. Here, the cross-layer resource allocation strategies for CRNs over the SDR testbeds were implemented through reconfigurable control and orchestration architecture. Further, for mission-critical applications, the authors in~\cite{ernesto2020mission} used the communication system with CRN using SDR. In this work, a cognitive protocol was implemented in the SDRs to facilitate mission-critical message transmissions. SDRs are also involved in anomaly detection in the networks, without too much dependence on the centralized architectures~\cite{katzef2021privacy}.

\subsubsection{Cognitive engine} 
The cognitive engine is considered to be the brain of cognitive radio since it maintains intelligence regarding sharing the spectrum efficiently. The cognitive engine is where the actual cognitive processes happen, which is the actual learning and decision-making for appropriate response based on the observed network behavior, specifically free spectrum~\cite{thomas2007cognitive}. In CRN, the cognitive engine makes the decision of spectrum validity and allocation. Cognitive engines are responsible for dynamic spectrum access through better situational awareness and intelligence~\cite{shah2021neural,jayaweera2018cognitive}. They are also involved in imparting security and privacy for a diversified range of applications. Further, Cognitive engines are also capable of imparting hybrid intelligence that could emphasize adaptive intelligence in CRN~\cite{olaleye2019hybrid}. 

\subsubsection{Cognition cycle}
A cognitive cycle involves sequences of stages accessed from the radio environment, towards spectrum sensing and spectrum characterization for effective means of making spectrum decisions. Moreover, the spectrum sensing stage could also detect the mobility of the PUs, and sensing the spectrum mobility could involve spectrum decision-making. Further, based on the channel capacity, spectrum-sharing tasks are carried out which support signal transmission back to the radio environment. This sequence of tasks forms a complete cognitive cycle, which is repetitive right from the connection established to the end of communication in a CRN. The impact of the cognitive cycles is largely observed in multimodal systems~\cite{komashinskiy2020introduction}, for network optimization tasks~\cite{oshima2018wireless}, collaborative sensing~\cite{ferreira2019tool}, and other associated CRN applications.

\subsection{Spectrum sharing techniques} 
As radio network possesses a spectrum of different frequencies, we can connect different wireless or mobile devices that operate on a wide range of frequencies for different types of communication. 5G services use both low and high-frequency bands and mostly in rural areas with a lower range of connectivity options, certain bands are left unused. Such unused spectrum could be shared with local communities, organizations, and service providers~\cite{agrawal2022spectrum}. Spectrum sharing and management techniques help to make potential usage of the whole spectrum.

Spectrum-sharing techniques of CRN are categorized based on the network architecture, allocation behavior, and access techniques

\subsubsection{Network architecture}
\begin{itemize}
    \item \textit{Centralized network:} In this scheme, a central controller is employed for controlling the access and spectrum allocation to the cognitive users. Here a base station could act as a  central controller and be responsible for making spectrum access decisions such as the transmission power for each device, and duration of the spectrum allocation. From the information collected by the controller on the demand of spectrum from the licensed users, optimal solutions on throughput, latency, and QoS could be estimated~\cite{rajkumari2018secure}. Such decisions made by the central controller are broadcasted and exchanged to the users, which leads to significant overhead in the transmission.
    
    \item \textit{Decentralized network:} In decentralized or distributed CRN, without the central controller the cognitive users could be able to share information with each other. This involves each CR transceiver of the users requiring more computation resources compared to the centralized solution. It enables them to independently make decisions locally on the spectrum access~\cite{sawant2019learning}. Further, the overhead in the communication will also be significantly reduced and due to the multi-hop nature of communication, occasionally the cognitive users might be considered relay stations 
\end{itemize}

\subsubsection{Spectrum allocation behaviour}
\begin{itemize}
    \item \textit{Cooperative spectrum sharing:} In this scheme, a common control channel or a centralized base station will be employed to establish cooperation among all the cognitive users. It reduces the sensing time and ensures maximum efficiency in the spectrum sharing among the users and also it improves the sensing accuracy with a commendable degree of fairness~\cite{tang2021cooperative}. However, the overhead, complexity, and energy consumption that are prevalent in cooperative spectrum sharing are playing a significant role in estimating the presence of primary users in the network.  Through clusters, the communication overhead could be significantly reduced in such a way that the central controller makes decisions based on the reports compiled from various cluster heads.

    Since spectrum sensing-based access cannot guarantee interference-free allocation to the users there has been a lot of progress done to develop database-assisted methods. Users and their spectrum allocations are registered in a database and the access is controlled based on requests and available resources. The most well-known approaches are TV White Space (TVWS) networks, Licensed Shared Access (LSA), and Citizens Broadband Radio Service (CBRS) systems~\cite{gurney2008geo, palola2017field, massaro2020will, hoyhtya2021licensed}. The database-assisted approach can be used e.g., to limit the number of users that can access the spectrum simultaneously in a limited geographical area. As a consequence the aggregated interference is controlled and primary users are protected.
    
    \item \textit{Non-cooperative spectrum sharing:} Here, unlike the cooperative scheme the cognitive users will not be involved in sharing any information among themselves. This mechanism will be helpful for sharing the spectrum among less number of users since the single user sensing information is sufficient enough to make decisions on the sharing of the channel~\cite{amjad2016evolutionary}. However, these non-cooperative mechanisms are prone to false alarms and will inevitably lead to interference events. Use of non-cooperative methods may lead to faster dynamic operations but also degrade the performance of the cognitive or primary users. 
    
\end{itemize}

\subsubsection{Spectrum access techniques}
\begin{itemize}
    \item \textit{Spectrum interweaves:} If the allotted spectrum for the primary user is not utilized at a particular sequence, time, or space, the spectrum interweave technique could be used as an opportunity for sharing the spectrum with other users~\cite{zhu2020machine}. This could be achieved by monitoring the activities of the primary users and by detecting inactive parts of the spectrum, the utilized spectrum band could be shared with the cognitive users. As soon as the primary users become active, the cognitive user will leave the spectrum and this interweave strategy is well exploited by frequency, time, or orthogonal division multiplexing for wireless communication systems. 
    
    \item \textit{Spectrum underlay:} The spectrum underlay approach can achieve concurrency in the transmission of the primary users with the cognitive users, provided the transmission power of the primary user should be high to avoid interference with the cognitive users. The tolerance level of interference is set by the receiver of primary users considering the interference temperature without affecting the normal operations~\cite{mohammadi2020resource}. The support for the spectrum underlay scheme could be rendered through ultra-wideband or CDMA technology. Unlike the spectrum interweave technique, the underlay approach avoids the need for a spectrum hole, however, it requires robust mechanisms for  avoiding interference between cognitive and primary users,

    \item \textit{Spectrum overlay:}    The spectrum overlay technique focuses on the concurrent means of transmission between cognitive and primary users, and it helps to mitigate the interference through proficient pre-coding strategies. Despite being the most promising technique, the spectrum overlay demands strong cooperation among the knowledge of message signals between primary users~\cite{safwat2018strategic}. Further, the cognitive user could utilize the power of primary users to relay the information of primary users, which in turn enhances the SNR of the primary user. More robust coding techniques are in demand for a spectrum overlay strategy at the cognitively transmitted end to mitigate the interference at the receiver end.
\end{itemize}

\subsection{Security Attacks on CRN}
Attacks on CRN result due to unauthorized access to the spectrum. Its impact will be catastrophic if a minimal number of adversaries involved in performing minimum operations are engaged to cause maximum loss to the mobile stations in the network. Attacks on cognitive networks may result either in unacceptable interference to the licensed primary users or missed opportunities for secondary users. In what follows, we present the most predominant attacks across the different layers of CRN.

\subsubsection{Physical layer}
\begin{itemize}
    \item \textit{Primary user emulation attack:} In this attack, the malicious intruder imitates primary user transmissions and prevents them from conveying and revealing their original information. It makes the secondary users believe that the channel is preoccupied~\cite{karimi2019smart}. Moreover, in this attack, since the intruder constantly listens to the channel for transmission opportunities, it consumes a lot of energy compared to conventional communication.
    
    \item \textit{Objective function attack:} Objective functions are one of the primary operation parameters of CRN, which are mostly used to optimize the operations of CRN. In the Objective function attack, the attacker tries to manipulate the objective function and takes control over the radio parameters~\cite{reshma2021security}. The attacks also manipulate the transmission rate parameters, so the computed results of the objective functions are biased toward the interest of the attacker.
    
    \item \textit{Jamming attack:} This kind of attack could lead to the degradation of the networks and even Denial of Service (DoS). Also, it involves the exploitation of free frequency bands. The attackers may target a single frequency (i.e., Spot jamming) or sweep across the range of frequencies (i.e., Sweep jamming), or may jam a range of frequencies in a single stretch (i.e., Barrage jamming). The attacks may also be done by a single jammer or collaborative set of jammers, who may extract more information on the channel and thereby reduce the throughput of the users~\cite{amjad2021ads}. The attackers could also continuously send jamming signals or alternate between the sleeping mode and attack mode. 
    
    \item \textit{Spectrum sensing data falsification:} Here, the malicious secondary users are involved in sending false sensing decisions. The attackers could be individuals or groups involved in manipulating or modifying the sensing results~\cite{marriwala2021authentication}. This significantly degrades the sensing accuracy of the network.

    \item \textit{Eavesdropping:} Through this unethical means of attack, the attackers could probe into the legitimate communication link~\cite{subbulakshmi2018mitigating}. Normally it occurs through jamming between the secondary pairs in the network. 
    
\end{itemize}

\subsubsection{Link layer}
\begin{itemize}
    \item \textit{Byzantine attack:} In a distributed network of computing nodes, it is a challenging task for the nodes to agree on a decision if some of the nodes act dishonestly. Byzantine attacks happen due to the failure of nodes to agree and execute common actions in a coordinated way~\cite{marchang2018detecting}. The messages could end up destroyed or made lost by the dishonest nodes in the network. 
    
    \item \textit{Common Control Channel Jamming:} Control channels in a network normally consume less amount of resources and are capable of sending control information across the network. An attacker could establish a DoS attack by jamming the control channels~\cite{manesh2018security}. Here, the attacker may not be involved in jamming the whole frequency band to hinder communication.      
    
    \item \textit{Common Control Channel Saturation:} In this kind of scenario, the network throughput of the channel is highly impacted, particularly when the number of users is high~\cite{sa2021role}. This network bottleneck impacts in wastage of network resources in terms of transmission and negotiation time constraints. 
\end{itemize}
    
\subsubsection{Network layer}
\begin{itemize}

    \item \textit{Ripple:} Ripple effects in the network layer often introduce false information about spectrum assignment in the network~\cite{sadkhan2019security}. Continuous trust management of the secondary users needs to be ensured to address these effects in the network.
    
    \item \textit{Sinkhole:} The malicious node referred to as a sinkhole node advertises an illegal route as the best route to the base station  or the sink node. The intermediate nodes across the path of the network from the sinkhole node then redirect their data to the sinkhole node upon receiving the broadcast illegal route rated as the best route~\cite{aslam2021sixth}. In addition, it may also be involved in modifying the data before relaying them to the base station or the sink node.
    
    \item \textit{Sybil:} With multiple incognito identities the attacker influences the network, which leads to restricted access to the primary users over the channel~\cite{rathee2020handoff}. The legitimate identification of a node could be exploited in the Sybil attack to create more identities and behaves as distinct nodes in the network.
    
    \item \textit{Wormhole:} Here the malicious nodes transmit the message over the low idle link and tunnel them to the other malicious node at the other end of the network~\cite{sadkhan2019security}. The attacker could modify the packets and make the data flow through the malicious nodes.
\end{itemize}

\subsubsection{Transport layer}
\begin{itemize}
    \item \textit{Key depletion:} During key depletion attacks, it increases the round trip times, and frequent re-transmissions of the data increase key usage~\cite{mishra2019cognitive}. This attack initiates a larger number of TCP sessions associated with a cryptographic key. It increases the chance of key duplication and aids the attacker to breach the system. 

\end{itemize}
 
\subsubsection{Application layer}   
\begin{itemize}
     \item   \textit{CR virus:} The bloom of artificial intelligence (AI) has introduced the cognitive radio (CR) virus, which is also called an AI virus. It is a self-propagating virus, which exploits the CRN in the application layer by taking advantage of the weakness in the characteristics of the network and the behavior of the users in the network. The impact of this virus increases exponentially with an increase in network size~\cite{sadkhan2019security}. Moreover, its spread on dynamic networks is reported to be higher than the static networks. Also, with the increase in spectrum channel resources, the speed of spread of the CR virus is observed to be higher.
     
     \item   \textit{Policy attacks:} In the policy attacks the dynamic changes in the policies of the radio are exploited by the attackers in terms of not allowing it to be updated or the attacker modifies the policy with false information~\cite{jararweh2018anomaly}. Based on the imperative features of cognitive wireless sensor networks, policy attacks are classified as excuse attacks and newbie-picking attacks. In excuse attacks, the network policy tends to be over-generous to the attackers and a malicious node may claim rights on the sharing of spectrum resources of crashed or damaged nodes. In the case of newbie-picking attacks, the new nodes before consuming any shared resource, pay their dues by requesting the network resources for a few periods. It leads to a veteran node moving from the newbie node, which may lead to a lack of information sharing in the application layer of CRN.
     
\end{itemize}

\subsubsection{Cross layer}
\begin{itemize}
     \item   \textit{Routing information jamming:} During the exchange of routing information between two nodes in an SDN, an internal malicious node targets the destination node and stops the exchange of data packets by simulating spectrum handoff signals~\cite{murmu2020security}.
     
     \item   \textit{Small Back-off window:} By the decrease of the back-off window of the node, they are prone to small back-off window attacks. In this attack the reduction in back-off window size makes the node to be malicious~\cite{asati2018rmdd}. This malicious may affect the data transmission and reception among other nodes in the network.  
     
     \item   \textit{Jellyfish:} Jellyfish attacks in SDN are involved in creating fake responses by occupying a closed loop during the attack~\cite{sadkhan2019security}. During the cognitive communication in the network, these fake responses may reflect the loss of data and delayed responses from other nodes in the network.

     \item   \textit{Lion:} The Lion attacks are triggered through the primary user emulation attacks in the physical layer, which causes TCP distortion. Logical connections are established by that distorted TCP and it will be engaged in continuous data packet transmission~\cite{sadkhan2019security}. During timeouts of packets, it tries for retransmission, which consumes double the amount of time and also leads to loss and delay of packets due to back-off. 
\end{itemize}

These challenges need solutions for their use in future communication networks, such as 6G. In the following section, we discuss how DL approaches provide viable solutions to these and other challenges followed by the applications of DL for important goals of CRN.

\subsection{Challenges in CRN}

\subsubsection{Interference mitigation}
Interference mitigation in a CRN requires a multifaceted approach encompassing spectrum sensing, resource allocation, and power control mechanisms. Adjacent-channel interference is observed when a signal from one channel intrudes into neighboring channels due to inadequate filtering. In contrast, co-channel interference takes place when several transmissions share the same channel, resulting in signal overlap and the possibility of degradation. To tackle these challenges, cognitive radios must detect unoccupied spectrum bands and anticipate possible interference by analyzing real-time spectrum utilization and environmental factors.

Enhancing system capacity while addressing interference necessitates sophisticated algorithms that adaptively optimize spectrum efficiency and interference management. Estimating the interference index is crucial in the design of CRNs, as it quantitatively assesses the interference levels experienced between secondary and primary users. The authors of \cite{ranjan2020interference} created a distributed greedy algorithm that compares the channel leakage ratio and interference index. This provides a scalable way to reduce interference in dense networks. This method significantly minimizes interference through optimizing the channel selection process, all the while enhancing spectrum utilization. In high-demand scenarios, such as natural disasters or emergencies, the network experiences significant traffic fluctuations, necessitating more robust and adaptive interference management techniques. Recent investigations have focused on applying ML and DL models for real-time interference prediction and dynamic spectrum management. 

\subsubsection{Exchange of control messages}
The network nodes exchange many control messages to coordinate CRN tasks like spectrum sensing, routing, group formation, and medium access. In~\cite{yadav2018approximating}, the authors devised a protocol involving one transceiver for exchanging control messages and another transceiver for data communication in a CRN network. Here, the protocol enables energy saving the CRN by switching between the sleep and active modes on detection of the PUs. Additionally, the exchange of control messages enables packet transmission control based on the network traffic. However, challenges are prevalent in exchanging the control messages on larger traffic and more nodes in the CRN. 

\subsubsection{Accuracy in the sensed spectrum}
Due to the scattered natures of cognitive radio devices across different locations, the cooperation among those nodes to operate with improved accuracy in spectrum sensing often remains a challenging issue. Further, the accuracy also depends on the number of channels and states in addition to the geographical locations of the nodes in the CRN~\cite{khaf2021partially}. Moreover, the correlation of the measured spectrum performed using efficient computation must be robust to noises with enhanced prediction accuracy. However, with the increasing number of users and interference in the network, the challenges in sustaining the accuracy in the sensed spectrum need to be addressed with advanced learning techniques.

\subsubsection{Seamless handover of spectrum}
The spectrum handover happens when the higher priority licensed user demands its channel over the SU. Seamless spectrum handover ensures a smooth transition of the spectrum from the SUs to the licensed user. Irrespective of the priority involved, the quick transition or handing over of the spectrum is quite challenging. Works on fuzzy logic techniques were used in research~\cite{anand2018fuzzy}, for spectrum handover considering the impact of power, channel reservation, and negotiation-based approaches. However, efficient decision-making to overcome the delay in handover and further optimization of power consumption, and illegal handover needs to be addressed by the research community.

\subsubsection{Optimal channel assignment}
In order to enhance the throughput, the SUs most often utilize multiple channels simultaneously. Also, efficient means of spectrum assignment is a key factor to realize dynamic spectrum access in CRN. Optimization is such a multi-channel assignment that aims to improve the throughput, selection capability, and channel assignment schemes. In~\cite{wei2018fair}, the authors addressed  the fairness issues on the SUs using a large number of idle data channels, by implementing a fair multi-channel assignment strategy. It ensures a considerable tradeoff between fairness and throughput. However, challenges do persist in imparting reliable transmission and addressing the issues in clustered multichannel CRN.

\subsubsection{Coordination between secondary users}
Throughput maximization and optimized energy consumption in a CRN could be positively enabled through strong coordination among the SUs in the network. Fusion centers are often responsible for establishing coordination among the secondary users, by collecting and processing the information from the users. However, in such centralized CRN, they are prone to the intrusion of malicious nodes, which could significantly reduce the performance of the networks and lack of coordination among the SUs. In~\cite{marvasti2019detection}, the authors addressed the spectrum sensing data falsification attack on one of the cooperative spectrum sensing scenarios. However, the lack of intelligence in decision-making on the attacks demands more robust learning-based approaches from the research community.

\subsubsection{Standardization of CRN}
CRN has seen a notable increase in standardization initiatives, especially concerning spectrum management and communication protocols. The goal of these projects is to create consistent frameworks that make it easier for cognitive radio networks to share and manage spectrum resources efficiently in a variety of settings \cite{khattab2015standardization}. As Patel et al.~\cite{patel2016achievable} points out, the IEEE 1900.5 standard is a big step forward in policy-based radio management because it makes dynamic spectrum access more efficient. Furthermore, the progress in 5G and B5G technologies has significantly heightened the necessity for CRN standardization, given the increasing importance of seamless communication between primary and secondary users. This standard is necessary to make interoperability easier, reduce interference, and improve resource allocation in cognitive radio networks. Such initiatives have created a base for reliable communication in complex wireless settings.

\subsubsection{Security awareness on the spectrum}
Security attacks on a network affect the reliability of the data communicated and particularly in the CRN, security awareness on the spectrum is of larger demand. By proper evaluation of the threats in the network, security awareness could be enhanced. Further appropriate location identification of the threats, decision-making capability, and robust security management features are also in demand. Bany et al.~\cite{bany2021intelligent}, developed an intelligent secure routing scheme for the CRN, considering in-band dynamic access networks, and ensures provides better spectrum management. However, the probability of success, full-duplex communication scenarios consideration, enhanced packet delivery ratio, and complete mitigation of the jamming effects still need attention.

\subsection{AI/ML/DL Approaches} 

AI modeling, training, simulation, testing, and deployment. Data-driven AI architectures are used in a wide range of wireless communication, particularly for CRN.

\subsubsection{ML Architectures}
\textcolor{black}{There are specific design rules that every ML model used in CRN applications must follow. These rules cover preprocessing methods, hyperparameter optimization, dataset selection, and training strategies for the models\cite{hlophe2021ai}. In CRN, these factors significantly impact how well they work, especially regarding spectrum sensing, resource allocation, and reducing interference.
}

\textcolor{black}{The preprocessing phase is essential in preparing input data for the learning models. To improve the spectrum data quality for CRN applications, methods like normalization, feature extraction, and noise filtering are likely to be used \cite{almuqren2024optimal}. For instance, raw spectrum data acquired via sensors often requires preprocessing to be converted into frequency or time-frequency domain representations. This can be achieved through techniques such as Short-Time Fourier Transform (STFT) or wavelet transforms, which subsequently serve as inputs for the learning models~\cite{pashaian2024speech}.
}

\textcolor{black}{An appropriate dataset is critical for effectively training deep learning models within CRN. In real-life situations, CRN datasets include spectrum occupancy data gathered from sensors working in real-time settings \cite{zhang2023pixel}. It is common for supervised learning tasks like spectrum sensing or modulation classification to use labeled datasets. These could come from RadioML, or software-defined radios (SDRs)~\cite{jagannath2022design}. The investigation has included an examination of the utilization of publicly available datasets alongside custom-collected CRN datasets across a range of studies, highlighting their distinct characteristics and the challenges they present.
}

\subsubsection{DL Architectures}
DL is a subclass of supervised learning approaches where we let the multilayer network perform the training and feature extraction tasks. DL enables better wireless communication services, particularly in data preparation. The CRN architectures require mobile stations, base stations or access points, and core backbone networks as their primary requirements. Based on these three components involved in the CRN, their network based on DL architectures may fall into any of the following kinds.

\textbf{Network architectures:}
Based on the connection topology involved in establishing the communication between the primary and secondary users with the access points, the backbone network architectures in CRN can be infrastructure-based, Ad-hoc based, or Mesh-based architectures. 
\begin{itemize}
     \item \textit{Infrastructure-based architecture:} In the CRN architecture with an infrastructure-based setup, the mobile stations access the base stations in a single hop. Even though the mobile stations are under the transmission range of the same base stations, communication is established between two users only through the base stations. The base stations may run multiple communication protocols to manage the diversified demands of mobile users.
	
     \item  \textit{Ad-hoc based architecture:} In the ad-hoc architecture implementation for the CRN, the network is set by on the fly without any significant support for the infrastructure. If other nearby users recognize mobile users, they are linked through communication strategies or protocols and form an ad-hoc network. However, the CRN nodes also uses other conventional wireless standards such as blacktooth and WiFi for effective spectrum occupancy dynamically on the spectrum holes. 
	
     \item \textit{Mesh-based architecture:} Mesh architectures integrate the best features from the infrastructure-based network and the ad-hoc architectures, enabling wireless connection between the access points. In this hybrid network, mobile stations can access the base stations directly or use multi-hop relay nodes based on their cognitive radio capability.
\end{itemize}

\textcolor{black}{\textbf{DL architectures for spectrum awareness:} Based on one of the crucial application demands of CRN, the role of deep learning in spectrum awareness has made a significant impact.}
\textcolor{black}{
\begin{itemize}
    \item \textit{Enhanced Performance in Complex Environments:} Deep learning models, like CNNs and RNNs, are better at finding complex patterns and reducing channel distortions than older methods like least squares (LS) and minimum mean square error (MMSE) in environments that are not linear, have a lot of noise, and fade quickly.
    \item \textit{Feature Extraction and Estimation: }DL models can directly take features from raw or already processed signal data to estimate channel state information (CSI). This cuts down on estimation errors and makes better use of the spectrum. Generative Adversarial Networks (GANs) are also used to simulate noisy environments. This gives realistic data for training and makes it possible to handle low SNR conditions robustly.
    \item     \textit{Improved Equalization and Efficiency: }DL-based equalization methods learn the opposite of channel distortions and change with the conditions. This leads to better signal recovery, higher throughput, and less need for frequent channel measurements, which makes CRN operations faster and more efficient.
\end{itemize}
}

\textcolor{black}{When using an RNN model to predict how the spectrum will be used in the future, choosing the learning rate and the number of hidden units is very important for how well the model makes accurate predictions~\cite{goyal2021deep}.
During the training phase, stochastic gradient descent (SGD) and backpropagation methods are often used to make deep learning models better \cite{masiha2022stochastic}. Different metrics, such as accuracy, precision, recall, and confusion matrices, are used to judge deep learning models in cognitive radio networks. These are especially useful for classification tasks like modulation recognition. We also showed examples of cross-validation methods to avoid overfitting and ensure that the results apply to a wide range of CRN environments.
}

\textcolor{black}{Depending on specific application requirements, various deep learning architectures are utilized within cognitive radio networks. CNNs have become very popular in spectrum sensing because they are good at working with spectrograms and other types of high-dimensional input data~\cite{mirza2024residual}. On the other hand, RNNs and LSTM networks are often used for tasks that need to analyze time, like predicting how spectrum will be used in the future~\cite{hlophe2021ai}. The discussion has encompassed utilizing hybrid models integrating CNN and RNN architectures for multi-task learning within CRN applications. Furthermore, attention mechanisms are used to improve model performance by focusing on essential input data components.
}
\subsection{The Need of DL in CRN}
DL can not only complete the applications in the field of CRN but also effectively could solve optimized usage of resources and enhance the security solution in CRN. Below we summarize the benefits of introducing DL in CRN from four prospects: Security, Computation, Decision modeling, and power control.

\begin{enumerate}
    \item \textit{Security:} CRN technology has expanded rapidly in recent years, but despite all its benefits and wide prospects of these networks, CRN is particularly vulnerable to security issues. They are associated with a larger number of nodes and their associated hardware and software units which increases their threats towards attacks and puts them at higher risks. Accordingly, CRN nodes with public interfaces are exposed to higher risk levels~\cite{aslam2021sixth}. Commonly, the attackers are involved in device tampering, phishing, and malware injection in the CRN nodes. DL models have evolved and laid the foundations for addressing the security issues in CRN~\cite{srinivasan2018ai}. 
    
    \item \textit{Computation:} DL models drive to provide reduced computational complexity and computation timings for CRN. They are also suited for delay-sensitive applications and could be extended for future wireless services and technologies as well. Furthermore, for security enhancement in the network, DL models could reliably classify the attackers and help to make systematic decisions for ensuring the intended transmission of vital information in the network~\cite{lee2021deep}. 
    
    \item \textit{Decision Modelling:} Sustained performance on correct transmission decisions with the support of DL models may fool the attackers to make false predictions. Moreover, control schemes based on the DL model supporting prediction strategy could save energy in the network and make a better prediction of the traffic load. Spectral decision models based on DL techniques assist to provide better performance through feature extraction done by the classifiers at different levels of network traffic conditions~\cite{giral2021spectral}. This scheme also enhances the performance in terms of providing better handoffs, reduced delay and optimized usage of bandwidth, and considerable throughput. 
    
    \item \textit{Power Control:}The matrices evaluated through DL techniques could suggest and classify the better channel and prioritize the data packets through those paths. Studies have also suggested that DL models could regulate the interference of primary users and distribute the strategy for transmission power control of secondary users~\cite{lee2021deep}. This ensures a near-optimal spectral efficiency and distributed power control strategy for CRN.
\end{enumerate}

\section{DL-based CRN}
\label{sec:DLforCRN}
In this section, a brief introduction to DL as well as to some of the major Deep Neural Network (DNN) architectures for future CRN, is presented. The objective of this section is to familiarize the readers with the fundamentals of DL techniques and their role in CRN applications. From the relevant references provided in each sub-sections, further details about the technologies can be explored. 


Among the several layers present in a  neural network, the first layer will be the input layer, which takes features as input from the given data and learns from these features. The intermediate layers are known as hidden layers, whose role is to process and merge input data. The final layer is the output layer, which is used for observing the outcome by predicting the value depending on the performance of the model. If the model is subjected to intensive training on a large volume of data very well, then the output categorization and classification will be more accurate. The difference between a neural network and DL is the number of hidden layers. If the number of hidden layers or the depth of hidden layers is increased for a neural network, it gets transformed into a deep neural network~\cite{liu2018deep,ahad2016neural}. DL algorithms are used in weather forecasting, computer vision, and smart-grid prediction due to high accuracy as well as in CRN~\cite{lee2018resource}.

\subsection{What type of DL algorithms are used in CRN?}

In CRNs, various types of DL neural networks are used to solve complex issues. These include Convolutional Neural Network (CNN) \cite{zheng2020spectrum,lee2017deep,merchant2018deep}, primarily meant for spectrum sensing to identify the presence of primary users in available frequency bands; Recurrent Neural Network (RNN) \cite{lopez2019primary}, focused towards predicting the availability of frequency bands over time; Long Short-Term Memory (LSTM) \cite{yu2018spectrum} and Auto Encoders, used for dimensionality reduction and anomaly detection in CRN signal patterns. Deep Stacking Network (DSN) and Genetic Adversarial Network (GAN) \cite{shi2018adversarial} are used for decision-making in dynamic spectrum allocation and for optimizing the transmission parameters in CRNs. Additionally, Deep Belief Networks (DBNs) are targeted for feature extraction and classification of the radio environment in CRNs.

DL algorithms have shown significant promise in addressing the complex challenges faced by CRNs. They enable data-driven decisions, improve network efficiency, and maximize spectrum utilization. Ongoing research in this field is expected to result in further advancements and optimizations to enhance the performance and capabilities of CRNs.

\subsection{How DL helps in CRN?}

With deep learning models, it is possible to make intelligent decisions such as spectrum selection, channel allocation, power control, and handover in real time based on various inputs such as channel conditions, interference levels, and user demands. Additionally, they can accurately detect signals in a specific frequency band and optimize the performance of CRN from the network management perspective ~\cite{li2020deep1}. For example, DRL can be used to dynamically adjust network parameters like transmission power and channel assignment to ensure efficient network utilization and improved user experience. Furthermore, deep learning models can be trained to classify different types of interference and take appropriate actions to reduce or eliminate them for interference mitigation in CRN.

Various types of CRN problems have been solved using deep learning algorithms. For instance, Wang \textit{et al.} in \cite{wang2019data} used CNNs for modulation recognition, while CNNs were used for symbol rate evaluation in CRN in \cite{merchant2018deep}. The frequency and bandwidth estimation problem in CRN was solved using a DNN in \cite{huang2018deep}. CNNs were also used for spectrum classification and sensing in \cite{lee_deep_2017} and RF fingerprinting in \cite{merchant2018deep}. CNNs are very effective for automated modulation recognition in CRN, with Rectified Linear Unit (ReLU) being the most commonly used activation function. Moreover, CNNs are useful for efficient resource management in wireless network systems \cite{zhang2019deep}.

\subsection{DL effect on spectrum sensing?} 
 
Spectrum sensing plays a crucial role in CRNs. In a recent study \cite{zheng2020spectrum}, the authors addressed spectrum sensing as a classification problem and proposed a DL-based solution. To mitigate the impact of noise power uncertainty, they normalized the received signal power and trained a CNN model with various types of signals and noise data. This approach enabled the model to perform well in detecting new, untrained signals. Furthermore, they utilized transfer learning techniques to enhance the efficiency of real-time signals. Simulation results showed that their method outperformed conventional spectrum sensing techniques such as the maximum-minimum eigenvalue ratio-based and frequency domain entropy-based methods.

Accurately detecting the primary user signal is critical for secondary users' safe utilization of idle licensed spectrum in CRNs. Traditional energy detectors are commonly used for blind signal detection, but they are limited by the well-known signal-to-noise ratio (SNR)-wall due to noise uncertainty. In \cite{gao2019deep}, the authors proposed DetectNet, a new DL-based signal detector that leverages the inherent structural information of modulated signals. DetectNet combines two DL algorithms, CNN and LSTM, for signal classification. Their results showed that using DetectNet significantly improved the efficiency of traditional energy detectors.

Similarly, Solanki et al. \cite{solanki2021deep} presented a DL-based model called DLSenseNet, which incorporates structural information from modulated signals to achieve high detection accuracy for spectrum sensing. The model was tested on the RadioML2016.10b dataset and demonstrated excellent performance. Chen et al. \cite{chen2020deep} proposed a collaborative spectrum sensing algorithm that utilizes distributed secondary users to acquire sensing samples and avoid fading and shadow effects. They proposed a DL-based detection framework, incorporating a CNN with the sample covariance matrix as the test statistic, called the CSSCNN algorithm. The authors also evaluated the algorithm's robustness against SNR and noise instability and demonstrated significant improvement in detection efficiency in complex scenarios. Xie et al. \cite{xie2020deep} proposed a DL-based model named CNN-LSTM for spectrum sensing. The model utilizes a CNN to extract energy-correlation features from the covariance matrices of sensing data and then uses LSTM to learn the primary user activity pattern. This approach enhances the probability of detection, even in scenarios with noise uncertainty. The authors demonstrated the superiority of the CNN-LSTM detector through ample simulations in both noisy and noiseless environments.

\textit{Links in CRN:}
In CRNs, links are crucial for ensuring the quality of service (QoS) for secondary users. The strength of the connection between the transmitter and receiver can be predicted before transmission, which helps in determining the QoS of secondary users. Both primary and secondary users rely on links with nearby base stations for effective communication with minimal delay~\cite{hlophe2019spectrum}. DNN models can learn from raw signal samples without requiring feature engineering, making them a promising solution for link quality estimation in CRN. CNNs and RNNs have been applied to link quality estimation by processing various input features such as raw signal samples, SNR, and channel state information (CSI) to predict link quality. In cognitive radio networks, links are crucial for ensuring the quality of service for secondary users. The strength of the connection between the transmitter and receiver can be predicted before transmission, which helps in determining the QoS of secondary users. Both primary and secondary users rely on links with nearby base stations for effective communication with minimal delay~\cite{hlophe2019spectrum}. DNN models can learn from raw signal samples without requiring feature engineering, making them a promising solution for link quality estimation in CRN. CNNs and RNNs have been applied to link quality estimation by processing various input features such as raw signal samples, SNR, and CSI to predict link quality.

\textit{IP Mobility Management in CRN:}
Deep learning techniques have been used to address mobility management challenges in CRN. For example, in~\cite{fourati2021survey}, a DRL model is proposed to optimize the handover decision in CRN. The DRL agent learns to make handover decisions based on channel quality, spectrum availability, and user mobility patterns. The authors demonstrated that their proposed DRL model outperformed traditional handover decision algorithms. Another approach to mobility management in CRN using deep learning is through CNNs and long short-term memory (LSTM) networks. In~\cite{zheng2020spectrum}, the authors proposed a CNN-LSTM model for predicting the optimal access points for mobile nodes. The model takes input features such as channel quality, spectrum availability, and user mobility patterns, predicting the optimal access point for the mobile node to connect. The authors demonstrated that their proposed CNN-LSTM model outperformed traditional handover decision algorithms and achieved higher accuracy in predicting the optimal access points.

\subsection{Federated Learning in CRN}
Federated Learning (FL) is a distributed learning technique that enables effective machine learning across multiple devices or computational nodes while maintaining server and user data privacy~\cite{shahraki2023machine}. Its primary purpose is to overcome the challenge of isolated data silos. FL is now being used in various fields such as medicine \cite{boughorbel2019federated}, CRN, and wireless communications \cite{wang2021federated}.

\subsubsection{The applications of Federated Learning}

In wireless communication, resource allocation is often done using FL. Recently, Samarakoonin et al.\cite{samarakoon2019distributed} proposed a solution for combining resource allocation and power control in ultra-reliable low-latency communication (URLLC) for vehicular networks. They developed an FL-based estimation approach to determine the length of a vehicle queue's tail distribution, and their simulations showed that FL is as effective as centralized solutions.
In a similar vein, M. Chen et al.\cite{chen2020joint} proposed an FL-based joint user selection and resource allocation scheme for wireless networks, considering realistic scenarios that account for factors such as packet errors and wireless resource availability. Unlike the previous study, their approach takes into account wireless factors, making it more practical. Moreover, Wang et al.\cite{wang2021federated} introduced an FL-based approach for distributed Automatic Modulation Classification (AMC) scenarios. They proposed a distributed AMC solution that trains FL models on each client rather than a centralized server to minimize the risk of data leakage. Their solution minimizes performance degradation while providing the advantages of FL-based AMC techniques.

\subsection{Impact of Federated Learning in CRN}
In the realm of Industry 4.0 scenarios, data-driven cognitive computing (D2C) faces significant challenges despite its numerous benefits. Privacy leakage, low efficiency, and a lack of incentives are among the most significant roadblocks to D2C adoption in smart manufacturing \cite{gu2019privacy}. The low efficiency is caused by the large volume of data generated by various Industry 4.0 devices \cite{pokhrel2020compound}. Data curators are hesitant to contribute their data to cognitive servers or service providers due to the privacy risks involved, which reduces their motivation to share data \cite{zhu2017cognitive}. These three challenges make it difficult to ensure accuracy in D2C adoption.

Authors in \cite{qu2020blockchained} proposed federated learning as a paradigm for cognitive computer learning. With federated learning, the same model is used on both the central server and the local servers at the same time. Only model updates are sent from local servers to the central server during the training process. Because of this, the privacy of the local data remains protected. In addition, the volume of model updates is substantially smaller than the actual data. It's safe to say that efficiency is also ensured in this manner.

Shachi et al. proposed the Spatio-temporal system model of cooperative spectrum sensing scenario \cite{shachi2020convolutional}, for which a convolutional neural network (CNN) is trained to achieve spectrum sensing. When compared to traditional techniques, their scheme achieves a higher detection probability. However, the algorithm sends the training data directly to FC, which consumes a lot of communication resources and reduces the efficiency of training and testing. Motivated by this, authors in~\cite{chen2021federated} provided the federated learning framework to cooperative spectrum sensing (CSS) and proposed a federated learning-based spectrum sensing method called federated learning-based spectrum sensing (FLSS). This architecture minimizes communication costs between FC and SUs and enhances model performance.

Since the federated learning framework uses the model parameters rather than training data, it consumes far less communication between FC and SUs than the current deep collaborative sensing approach. As a further benefit, FLSS trains the model in each SU instead of just the FC, which maximizes processing power and improves training efficiency.

\begin{figure}[h] \centering        
    \centerline{\includegraphics[width=0.50\textwidth]{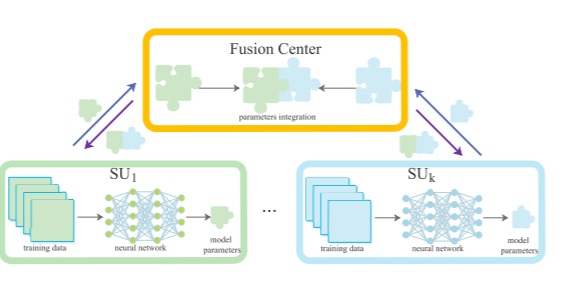}}
    \caption{FL based spectrum sensing: Off line training \cite{chen2021federated}}
    \label{fig:FL1}
\end{figure}

\begin{figure}[h]        
    \centerline{\includegraphics[width=0.50\textwidth]{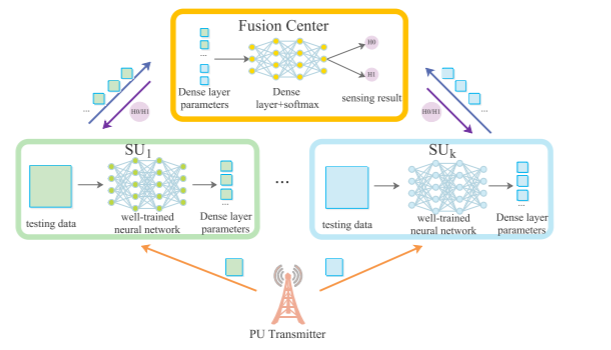}}
    \caption{FL based spectrum sensing: online line detection \cite{chen2021federated}}
    \label{fig:fl2}
\end{figure}

According to Fig. \ref{fig:FL1} and Fig. \ref{fig:fl2}, the spectrum sensing method is broken down into two parts: offline training and online detection \cite{chen2021federated}.

\section{DL for Spectrum Awareness and Modulation Recognition}
\label{sec:spectrum}
In this section, inspired by the aforementioned applications of DL in CRN, we present a comprehensive overview of the key research highlights that must be pursued for the practical deployment of spectrum monitoring, modulation recognition, and transmission in CRN. For each of the mentioned applications, we outline the overview of recent results from the literature addressing the mentioned challenges of CRN in the aforementioned sections. 

\subsection{Spectrum monitoring}

In~\cite{lee_deep_2017}, by considering spatial and spectrum correlation, better performance is achieved in cooperative spectrum sensing by incorporating CNN. Here, the authors established cooperation among the secondary users for the detection of primary user bands. The authors highlight the incorporation of deep cooperative sensing through CNN by combining the sensing outcomes from the individual secondary users. Further, the deep sensing strategy exhibits autonomous learning based on the environment and the individual sensing outcomes, which were observed for spatial and spectral correlation. 

For the implementation of spectrogram classification in CRN, thirteen different ML / DL techniques are analyzed by Lees et al.~\cite{lees_deep_2019} in the 3.5GHz band of radar detection. From the experimentation, it was observed that the three-layer CNN exhibited better performance with better trade-offs between the computational complexity and accuracy in classification. Further, the network was applied for extracting descriptive statistics from the full spectrogram library. However, from the techniques deployed using various ML and DL frameworks and comparisons made over classical methods, it was perceived that for radar detection over the range of 3.5GHz more sophisticated deep learning architectures are recommended.
 
Reinforcement learning with graph neural networks is employed for energy efficiency optimization at different scales of CRN and improved performance is observed~\cite{he_deep_2019}. Here, the authors focused on optimizing the spectrum sensing process and efficient energy utilization through combinatorial optimization over distributed cooperative sensing. Moreover, the deployment was evaluated over different ranges of network scales. However, remarkable breakthroughs are expected for application-specific energy efficiency optimization.

For carrying out Spatio-temporal sensing and signal classification, SVM techniques are observed to be robust to temporal detection of spectrum~\cite{awe_spatio-temporal_2018}. The method uses a feature realization strategy based on a beamformer for the enhancement of signal classification tasks by considering the operational scenario of both single and multiple primary users in the network. Furthermore, for solving the issues in  multiple state spectrum sensing,  error-correcting output codes were experimented with by the authors in ~\cite{awe_spatio-temporal_2018} by establishing alternate multiple independent models. However, irrespective of the better overall classification accuracy of the detectors, they need to experiment on robust conditions for the detection of spectrum holes in the CRN. 

In~\cite{han_spectrum_2017}, by  employing CNN better probability in detection is observed for spectrum sensing compared to cyclostationary feature detection. Here, the authors extracted and pre-processed the features of primary user signals and noise signals, which includes energy and cyclostationary features. Further, they are used as training input for CNN with the objective to detect the presence of primary users. Despite better detection probability, the signal-to-noise ratio enhancement for a diversified range of applications needs attention by incorporating more robust deep neural network-based spectrum sensing techniques.

Cooperative sensing of the spectrum is achieved using transfer learning by Do et al.~\cite{do_learning_2018}, through which enhanced the security level of the network with better data protection is observed. The authors investigated energy-efficient strategies for communications and employed secondary users to harvest energy, in which the decisions on spectrum sensing carried out by the secondary users are sent to the fusion center.  This was made feasible through CNN, in which the fusion center utilizes the data about the presence of primary users and spectrum sensing decisions to observe the active status of the users. Further, it was guaranteed to provide data encryption through private-key encryption, targeted for maximization of the network security level. However, the numerical evaluations on the configurations of the network need to be performed for more dynamic CRN settings.

Moreover, spectrum classification under colored noise environments is experimented with using CNN and better performance is projected~\cite{zheng_spectrum_2020}. The authors trained the model with signals and noise data to enable the model to be adaptable to new signals. Further, based on the extensive investigation of the normalized signals and noise power, it was observed that the deep learning-based classification outperforms two conventional spectrum sensing strategies. In specific, the deployment of transfer learning in this experimental setting shows an improved detection ratio. However, considering colored noises and real-world signals for validation of the superiority of the developed deep neural network could be adapted for CRN applications.

\subsubsection{Spectrum occupancy}
Spectrum occupancy aims at measuring the proportion of time when a certain frequency channel or frequency band is occupied in a given area ~\cite{7460899}. Here, the authors proposed a cautionary perspective  for expressing strong conclusions on the acquired data.  Spectrum measurement and analysis over spatial domain requires use of so-called interference maps or radio environment maps and before strong conclusions can be done the researchers should know how measured signals and frequency bands are actually designed to be used.  Further, with the spatial domain considerations, the improvement in the measurement accuracy was analyzed by considering different phases involved in the measurement process of the spectrum.  Accordingly, through this improved design, it is anticipated to obtain realistic spectrum occupancy data for taking accurate spectrum management decisions.
 
Ding \textit{et al.}~\cite{ding2017spectrum} presented a comprehensive survey on the spectrum interference, and spectrum occupancy statistics along with its sources in CRN. Further, the taxonomy of spectrum inference was introduced by considering time-frequency-space parameters. Besides, the authors also evaluated the efficiency of the interference with a focus on next-generation communications systems. Furthermore, the work discusses various applications of spectrum inference in CRN, along with some possible open research issues to alter challenges associated with CRN applications.

Distributed CRN with collaborative spectrum sensing, normally faces spectrum occupancy reconstruction. Singular value decomposition is used for solving the problem by minimizing the reconstruction error~\cite{hlophe2019spectrum}. Here, the spectrogram occupancy is addressed using deep neural networks operating with stochastic gradient descent and by solving the matrix factorization, correlations are done in distributed CRN. 

A spectrum prediction framework is developed using a DL approach in ~\cite{yu2018spectrum}. The authors used Long short-term memory and compared their performance over multi-layer perceptron networks on two datasets and observed the performance. The researchers applied the Taguchi method for determining the best fit configuration optimized for the neural network, and they analyzed the effects of the involved hyperparameters. In such an experimental setting, their novel approach could analyze and predict the performance at various frequency bands, and thereby determine and observe its stable performance with a classification perspective of  spectrum prediction.

For a multi-class classification problem, DL frameworks are adopted for observing spectral characterizations and prediction using superimposed radio frequency data under diversified scenarios of cognitive radio systems~\cite{omotere2018spectrum}. Their DL framework is trained with real traces of radio frequency data employed in different coexisting scenarios, collected from Universal Software Radio Peripheral-based testbeds. Their results demonstrate the feasibility of spectrum situation prediction, which is investigated in the complicated coexistence scenarios, that highlights the potential of the DL approach in this scenario. In conjunction with this process, the model is also capable of identifying the presence of signals at different signal levels from various sources of interference. 

With the focus towards a more realistic scenario, the 2D-long short-term memory model is used by exploiting the frequency and time correlation in the cognitive radio systems~\cite{aygul2020spectrum}. Therefore, the investigation results show improved performance with the consideration of computational complexity and accuracy in the identification of spectrum occupancy predictions. However, validations of the measurements over real-world spectrum still lead to being considered for the operation in a pre-defined range of frequency spectrum. Trying to solve this matter on several private uplink bands imposes trust issues, and costs and thus may lead to an even more expensive solution.

\begin{table*}[hbtp]
  \begin{center}
	\caption{\textcolor{black}{Contribution of DL Algorithms in CRN for Spectrum Sensing}}
	\label{table3}
	\begin{tabular}{|p{2.5cm}|p{2.5cm}|p{3cm}|p{3.5cm}|p{4cm}|}
	\hline
	\textbf{References} 
        & \textbf{DL Framework} 
        & \textbf{CRN Application} 
        & \textbf{\textcolor{black}{Datasets Used}}
        & \textbf{\textcolor{black}{Accuracy / Metrics}}
        \\ \hline
	  Lee \textit{et al.}~\cite{lee_deep_2017},
	  & CNN & Cooperative spectrum sensing & \textcolor{black}{Collected real-world spectrum data} & \textcolor{black}{Accuracy: 89\%, Better performance in spatial/spectrum correlation} \\
\hline
	  Lees \textit{et al.}~\cite{lees_deep_2019},
	  & Thirteen ML / DL techniques & Spectrogram classification & \textcolor{black}{3.5 GHz band radar detection dataset} & \textcolor{black}{Accuracy varies across techniques, ML outperforms traditional methods} \\
\hline
	  He \textit{et al.}~\cite{he_deep_2019},
	  & Reinforcement learning with GNN & Energy efficiency optimization & \textcolor{black}{Custom network simulation data} & \textcolor{black}{Energy efficiency improvement: 15\% compared to baseline} \\
\hline
	  Awe \textit{et al.}~\cite{awe_spatio-temporal_2018}, & SVM & Spatio-temporal sensing, signal classification & \textcolor{black}{Simulated spatio-temporal data} & \textcolor{black}{Robustness: High, Temporal accuracy: 88\%} \\
\hline
	  Han \textit{et al.}~\cite{han_spectrum_2017}, & CNN & Spectrum sensing & \textcolor{black}{Custom spectrum data} & \textcolor{black}{Accuracy: 92\%, Better detection probability compared to cyclostationary detection} \\
\hline
	  Do \textit{et al.}~\cite{do_learning_2018}, & Transfer learning & Cooperative sensing & \textcolor{black}{Cross-domain spectrum data} & \textcolor{black}{Security enhancement, Accuracy: 87\%} \\
\hline
	  Zheng \textit{et al.}~\cite{zheng_spectrum_2020}  & CNN & Classification under colored noise & \textcolor{black}{Simulated spectrum data under noisy conditions} & \textcolor{black}{Accuracy: 85\%, System performs better in noisy environments} \\

	   \hline
	\end{tabular}
  \end{center}
\end{table*}
\subsubsection{Spectrum sensing}

Cooperative spectrum sensing in CRN is carried out by Lee \textit{et al.}~\cite{lee2017deep} using CNN by considering environmental-specific cooperative spectrum sensing. Here, based on the environment-specific cooperative spectrum sensing, spatial and spectral correlations are considered for individual sensing. In this methodology of DL-based cooperative sensing, improved performance is guaranteed through the cooperation among multiple secondary users with primary users in the network. Further, a better improvement in the performance of cooperative spectrum sensing  is observed with a moderate range of training samples, which is guaranteed with the lower complexity of operations.

CNN-based DL model is used for spectrum sensing~\cite{xie2019activity}, in which primary user activity patterns are learned and detected using a completely data-driven approach, unlike the conventional signal-to-noise probability model or user activity model. Here, the current and historical sensing data are considered, which assists in assessing the activities of the primary user. The numerical results help to envisage the detection probability of the DL model and highlight its superiority over the conventional estimator-correlator and Markov model detectors.

The Support Vector Machine (SVM) approach is used for sensing the spectrum in multi-antenna CRN in~\cite{awe2018spatio}. In this approach, scenarios of multiple primary users are considered along with beamformer-aided features to enhance the SVM algorithm capability for classifying the signal under different user operating conditions. Further, the error-correcting output codes were considered for signal classification driven by multi-class SVM algorithms. This spatiotemporal spectrum sensing strategy in CRN provides a better probability of false alarms, overall detection, and classification accuracy. 

\subsubsection{End-to-End learning}
When Deep Neural Networks are used for training a complex system using a single learning model is referred to as End-to-end learning. It helps to learn the features automatically from the wireless signals and eliminates the demand for complex processing pipelines. End-to-end learning from the perspective of spectrum monitoring tasks is used for modulation recognition and interference detection using different representations based on amplitude, phase, and IQ data representation~\cite{kulin2018end}. This conceptual framework enables flexible implementation of signal classification and the authors assessed the accuracy variations for recognizing modulation and in the detection of signal interferences. 

End-to-end signal processing is also used for big data processing on transmitted wireless data for monitoring the spectrum and security aspects of CRN~\cite{zheng2018big}. Further, the authors explored the role of end-to-end signal processing based on DL frameworks in cyberspace security along with spectrum monitoring solutions. It also assists in provisioning distortionless compression of wireless data and unified representation of the signal features through DNNs. Channel occupancy is predicted using DL approaches for estimating the time slots for the channels~\cite{shenfield2020deep}. Here, the availability of the consecutive time slots is assessed and it predicts the underutilized spectrums, without interference over the occupied channels.

\subsubsection{Spectrogram classification}
Spectrograms are a visual representation of any signals that illustrate the amplitude of the corresponding frequency components of those signals over time. The authors in~\cite{li2020deep}, developed a CNN that utilizes a key point estimation approach for multitype signal detection and it recognizes the appropriate classes in the spectrum. It was involved in the classification of wideband spectrograms with local and border offset bounds. This approach abandons the candidate anchors and runs at a faster rate than conventional methods. Further, in one of the similar works reported in~\cite{huizing2019deep}, micro-doppler spectrograms are used for automatic target recognition applications, by exploiting the micro-doppler signature in the DL-based classification phase. Here, the experimentation was carried out for mini-UAVs by considering the properties of rotation of wings, rate of rotation, length of wings, and the number of wings in the UAVs. It provides a promising contribution to automatic target recognition applications in radars, considering its imaging principles and target scattering mechanisms.

\begin{table*}[hbtp]
  \begin{center}
	\caption{\textcolor{black}{Contribution of DL Algorithms in CRN for Spectrum Prediction}}
	\label{table4}
	\begin{tabular}{|p{2.5cm}|p{2.5cm}|p{3cm}|p{3.5cm}|p{4cm}|}
	\hline
	   \textbf{References}  & 
	   \textbf{DL framework} & 
	   \textbf{CRN Application} & 
          \textbf{\textcolor{black}{Datasets Used}} & 
          \textbf{\textcolor{black}{Accuracy / Metrics}} \\
	  \hline
	  Yu \textit{et al.}~\cite{yu_spectrum_2018} 
        & LSTM, Taguchi method 
        & Prediction of occupancy and quality of channel 
        & \textcolor{black}{Real-world wireless spectrum dataset }
        & \textcolor{black}{Accuracy: 90\%, Improved LSTM performance on real-world data with enhanced prediction accuracy} \\
\hline
	  Yu \textit{et al.}~\cite{yu_spectrum_2017} 
        & DL 
        & Aerospace communications 
        & \textcolor{black}{Synthetic aerospace data} 
        & \textcolor{black}{Prediction accuracy: 88\%, No a priori knowledge of primary user activities required} \\
\hline
	  Han \textit{et al.}~\cite{han_power_2020} 
        & Generative adversarial networks 
        & Power spectrum maps estimation 
        & \textcolor{black}{Simulated power spectrum data} 
        & \textcolor{black}{Accuracy: 92\%, GANs outperform conventional networks in estimation performance} \\
\hline
	  Supraja \textit{et al.}~\cite{supraja_optimized_2019} & Genetic algorithm 
        & Optimized neural network 
        & \textcolor{black}{Custom neural network optimization dataset} 
        & \textcolor{black}{Prediction accuracy: 91\%, Genetic algorithms provide better weight optimization for prediction tasks} \\
\hline
	\end{tabular}
  \end{center}
\end{table*}

\subsubsection{Spectrum Prediction}
Instantaneous information alone is not sufficient data in order to select the most suitable channels for cognitive transmission ~\cite{hoyhtya2011improving, hoyhtya2019database}. With clever use of historical data, one can estimate the collision rate and predict the desired channel selection. For instance, the unused channel for a longer period of time could be predicted through the rules applied based on stochastic ON-OFF and  partially deterministic patterns. Further, different prediction rules apply to stochastic and deterministic traffic patterns and the former allows only statistical predictions. With appropriate spectrum prediction strategies, unlicensed users can clearly increase their QoS and reduce the number of channel switches.

Using LSTM and Taguchi methods, prediction of occupancy and quality of the channel is carried out by Yu \textit{et al.}~\cite{yu_spectrum_2018}, and better performance of LSTM was predicted while operating on a couple of real-world spectrum datasets. It was observed that the prediction performance largely depends on the frequency bands of operations. Here, the authors explored the performance statistics of LSTM to be much better than the MLP network. With the best-optimized configuration of LSTM, the spectrum prediction problem could be more flexible in the prediction of the channel occupancy states. DL frameworks are also widely used in aerospace communications such as spectrum prediction carried out without prior knowledge of primary user activities~\cite{yu_spectrum_2017}.  Here, the analysis was carried out  through intensive simulations to predict the spectrum availability. This helps to enhance cognitive aerospace communications, by enabling the secondary users to access the unused licensed spectrum, which was owned by the primary users. 

The authors Han \textit{et al.} in~\cite{han_power_2020}, experimented on the radio resources to estimate the power spectrum maps. They implemented GANs for power spectrum maps estimation and from the results, accurate estimation in terms of performance is observed further, they were compared with the outcomes of conventional networks. Here, from the constructed model, initially, the estimation tasks were converted into image reconstruction tasks and followed by that, an image generator and discriminator are trained by the developed model. Optimization of neural networks for spectrum prediction was performed by Supraja \textit{et al.}~\cite{supraja_optimized_2019} and it provides the best weights for the network and  high prediction accuracy. This was experimented with using genetic algorithms to avoid the trap of locally optimal solutions. Further, by increasing the randomness and mutation functions, global optimal solutions were provided for alleviating the spectrum utilization challenges.

\subsubsection{Spectrum data poisoning}
Adversarial DL is used for managing spectrum data poisoning challenges in CRN~\cite{shi_spectrum_2018}. In this case, data poisoning attack on the spectrum is reduced and throughput is improved to a larger extent.

\subsubsection{User characterization}
For primary user characterization, LSTM and RNN are trained to enhance the improvement in prediction percentage compared to multilayer neural networks~\cite{lopez_primary_2019}. Deep Belief Networks are used to improve the accuracy in user agent classification, which are ensured with appropriate tests on Wifi channels~\cite{cui_deep_2015}.

\subsection{Modulation}
\subsubsection{Modulation classification}

\textcolor{black}{Modulation classification in CRNs poses a crucial challenge, as accurately detecting modulation schemes like QAM and PSK used by primary users is essential for efficient spectrum sensing and resource allocation. Researchers have developed various deep learning models for this purpose. CNNs are very good at automatic modulation classification because they can handle spectral data and pull out important features from raw in-phase and quadrature-phase signals. A CNN model from~\cite{wang2019data}, trained on the RadioML dataset, was able to recognize QAM and PSK signals 86\% of the time, even when the SNR was low. RNNs, especially LSTM networks, excel in analyzing time-series data. In CRNs, RNN-based approaches have been used to classify modulations by evaluating temporal signal characteristics. Sun et al. (2018)~\cite{sun2019using} reported an 88\% classification accuracy using RNN with synthetic spectrum data. GANs have been introduced to create synthetic spectrum data for training deep learning models, particularly when real-world data is limited. The methodology using GANs, as detailed by authors in~\cite{ayanoglu2022machine}, enhances training dataset diversity and model robustness under challenging channel conditions, achieving up to 91\% classification accuracy. Hybrid models that combine CNN and LSTM networks have been explored to utilize both spatial and temporal signal characteristics. As described in~\cite{ivanov2019hybrid}, these hybrid models perform better in environments with low SNR and fading, achieving a modulation classification accuracy of 90
}

Modulation classification with programmatic data augmentation is performed by Tang \textit{et al.}~\cite{tang_digital_2018} using Auxiliary classifier GAN. For automatic classification, deep belief networks are used in~\cite{mendis_deep_2017}, in which the low-complexity binarized network provides better accuracy. Autoencoders are configured for signal reconstruction and optimal classification in hybrid noise-resilient architectures under low SNR and fading conditions~\cite{ivanov_hybrid_2019}.

\textcolor{black}{Incorporating these wider range of deep learning models offers a deeper analysis of recent developments in automatic modulation classification for CRNs. This also highlights the effectiveness of various deep learning architectures, underlining the importance of selecting an appropriate model suited to the specific challenges of CRNs, such as noise, interference, and dynamic spectrum use.}

\subsubsection{Modulation recognition}

The CNNs are trained on different datasets by considering in-phase and quadrature-phase components for classifying QAM signals thereby driving towards automatic modulation recognition~\cite{wang_data-driven_2019}. However, the challenges in radio signal processing in the automatic modulation recognition are addressed by GAN neural networks more efficiently and it avoids non-convergence and improves the accuracy in recognition of the signals~\cite{li_generative_2018}.  By using naively learned features the efficacy of radio modulation classification in the complex temporal radio signals is done by employing CNN~\cite{oshea_convolutional_2016}. Hybrid learning is done in~\cite{arumugam_modulation_2017}, using the  side information and the higher-order moments with flexible time-space decompositions for modulation recognition.

\subsection{Transmission}

\subsubsection{Transmission scheduling}
For CR-based IoT applications, transmitter scheduling is performed using Deep-Q-learning, which is involved in addressing the packet transmission challenges by utilizing Q-learning with the minimum amount of resources~\cite{yeganeh2023novel}. Very recently, Salh et al.~\cite{salh2022smart} investigated on smart packet transmission scheduling mechanisms in cognitive IoT systems. The authors experimented with the system using a Generative Adversarial Network and Deep Distribution Q-Network (GAN-DDQN) for enhancing the packet transmission rate, which was achieved by reducing the spacing between the target and destination action-value particles. Moreover, they also emphasized the outcomes with reduced power consumption and minimized transmission dealy compared to conventional networks.

\subsubsection{Transmitter classification} 
It is highly essential for wireless services to be aware of the surrounding RF frequencies, in order to adapt their transmission and reception capabilities based on their demands. However, due to malicious attacks and spoofing, most of the techniques are ineffective for the faithful classification of transmission frequencies. In~\cite{roy2019rfal}, the authors used an adversarial learning framework for the classification of adversarial transmitters, thereby providing a quality solution for the detection and elimination of such transmitters. More specifically, the trusted and counterfeit transmitters were discriminated against through the discriminator model applied over the SDRs, which were based on universal software radio peripherals (USRP). Transmitter classification on CRN is done by Morin \textit{et al.}~\cite{morin_transmitter_2019}, in which supervised DL is used for reducing the channel biases and new datasets are created for future research.


\subsubsection{Space communication} The evolution of space communications has advanced towards being cognitive and begins to provide an improved range of spectrum efficiency and energy utilization. Murugan \textit{et al.}~\cite{murugan_efficient_2019} developed DL-based frameworks for enhancing communication between satellites and base station terminals. Here, the authors considered spectrum interference as a key issue for empowering the effectiveness of cognition in energy and range of communication.

Three-dimensional (3D) networks driven by 6G communication services will be dynamic, requiring sophisticated spectrum management that will use AI algorithms. As is envisioned in  ~\cite{9893104}, database-assisted technologies and AI-based optimized spot-beam settings could be used to ensure predictable QoS for all users.
\section{DL for Resource Allocations}
\label{sec:resource}

This section presents a comprehensive discussion of how the DL frameworks are used for the implementation of various ranges of resource allocation in CRN. In particular, it presents the message passing, multi-channel, and multi-efficiency allocation resources, including imparting the QoE in resource allocation tasks. Furthermore, the role of DL and resource management issues are discussed with a focus on key CRN applications.

\begin{table*}[!hbtp]
  \begin{center}
	\caption{\textcolor{black}{Contribution of DL Algorithms in CRN for Resource Allocation}}
	\label{table2}
	\begin{tabular}{|p{2.5cm}|p{2.5cm}|p{3cm}|p{3.5cm}|p{4cm}|}
	\hline
	\textbf{References} & 
	\textbf{DL framework} & 
	\textbf{CRN Application} & 
        \textbf{\textcolor{black}{Datasets Used}} & 
        \textbf{\textcolor{black}{Accuracy / Metrics}} \\
	\hline
	Liu \textit{et al.}~\cite{liu_deep_2018}  & Feed-forward neural network & Damped 3D message passing algorithm 
 & \textcolor{black}{Custom dataset of interference scenarios} 
 & \textcolor{black}{Accuracy: 88\%, Parameters optimized for interference reduction, Performance improved by 10\%} \\
\hline
	Liu \textit{et al.}~\cite{liu_multi-efficiency_2018} & DL & Reweighted message-passing algorithm 
& \textcolor{black}{Spectrum and energy efficiency dataset} 
& \textcolor{black}{Energy efficiency improvement: 12\%, Spectrum efficiency improvement: 15\%} \\
\hline

	Liu \textit{et al.}~\cite{liu_deep_2018-1} & RNN & energy efficient allocation for NOMA heterogeneous IoT 
& \textcolor{black}{IoT spectrum usage dataset} 
& \textcolor{black}{Accuracy: 90\%, Outperforms OFDM-based approaches in resource allocation} \\
\hline

	Lee \textit{et al.}~\cite{lee_resource_2018} & DNN 
& transmit power allocation to secondary users & \textcolor{black}{Real-world spectrum dataset} 
& \textcolor{black}{Spectral efficiency improvement: 18\%, Interference reduction: 8\%} \\
\hline

	Shah \textit{et al.}~\cite{shah-mohammadi_deep_2018}  & DRL & Deep Q- Network algorithm 
& \textcolor{black}{Simulated multi-user spectrum allocation data} 
& \textcolor{black}{Reduced number of iterations by 15\%, Improved resource allocation accuracy by 92\%} \\
\hline
	Mohamedou \textit{et al.}~\cite{mohamedou_dynamical_2016} & DNN based forcasting engine & Dynamic power allocation and spectrum sharing 
& \textcolor{black}{Real-time spectrum sharing dataset} 
& \textcolor{black}{Transmission power allocation improved by 10\%, Better resource utilization} \\
\hline
	\end{tabular}
  \end{center}
\end{table*}

The computational efficiency of resource allocation is improved by using a DL-based message passing algorithm~\cite{liu2018deep}. In particular, the optimal parameters of the DL network are learned using the developed backpropagation algorithm. 

Joint resource allocation with spectrum sensing along with power allocation is done using simultaneous wireless information and power transfer in Non-orthogonal multiple access cognitive networks~\cite{song2019joint}. For the Cyber-Physical-Social System using the Internet of Things technologies, resource allocations are carried out using the Jaya algorithm~\cite{luo2018resource}. With the popular usage of ML algorithms across diversified domains of research, they are also widely used for resource allocations in CRN. K-means ML algorithm is used for uplink resource allocation in multi-carrier grouping cognitive internet of things~\cite{liu2020uplink}. Also, for IoT applications spectrum allocations are performed using an effective multi-objective Optimization algorithm~\cite{han2018effective}. CRN for IoT is also treated with energy and spectrum efficient schemes~\cite{aslam2018energy}. A distributed networking framework is introduced with four layers with a hierarchical multi-agent system ~\cite{wang2019intelligent} for dynamic resource allocation in time-frequency space. 

By utilizing the Feed-forward neural network the parameters for optimizing the interference are done with the support of a damped 3D message passing algorithm and improved performance is observed~\cite{liu_deep_2018}. DL frameworks are configured with reweighted message-passing algorithms by  Liu \textit{et al.}~\cite{liu_multi-efficiency_2018}, and drastic improvements in the efficiency of energy and spectrum for primary and secondary users are observed. For NOMA heterogeneous IoT, energy-efficient resource allocations are done, and compared to OFDM usage of RNN provides better results~\cite{liu_deep_2018-1}. DNN is also used for enhancing the spectral efficiency of the secondary users with less interference by appropriate transmit power allocation to secondary users~\cite{lee_resource_2018}. Shah \textit{et al.}~\cite{shah-mohammadi_deep_2018}  utilized the Deep Q- Network algorithm with a reduced number of iterations in the learning phase without the support of the transfer learning part for effective resource sharing by the primary and secondary users.  A DNN-based forecasting engine is developed by Mohamedou \textit{et al.}~\cite{mohamedou_dynamical_2016}, which provides better transmission power allocation through the integration of dynamic power allocation and spectrum sharing components in the DNN. 

A few of the most prominent categories of resource allocation through DL techniques are elaborated as follows:

\subsection{Message Passing}
To establish better coordination in the sensing policies of the cognitive network, it is desirable to obtain them in a decentralized fashion. Message-passing frameworks enable such coordination with the speed in sensing and improved accuracy. A damped 3-dimensional message-passing algorithm is developed based on DL using a feed-forward neural network to learn the parameters of message passing for optimized resource allocation~\cite{liu2018deep}. To minimize the interference power, a reweighted message-passing algorithm is proposed using DL~\cite{liu2018multi}. Message passing frameworks are also implemented using CNN~\cite{wang2019data}, for improving the accuracy in recognition. Optimized resource allocation through message passing is implemented for a NOMA heterogeneous IoT using a DL-based scheme for providing low computational complexity and faster convergence~\cite{liu2018deep1}. Muti-agent based Deep Reinforcement Learning method is used for expensive computation and communication overhead through the message passing implemented through Q-network strategy~\cite{zhao2019deep}.

\subsection{Multi-efficiency based allocation}
A multi-efficiency-based scheme of resource allocation is performed using a DL-based resource allocation algorithm in cognitive networks for improving the spectrum efficiency of the users, improving the computing efficiency and energy efficiency of the system~\cite{liu2018multi}.

\subsection{Multi-channel underlay}

In multi-channel CRN, the interference of secondary users on the primary users is minimized to a larger extent using deep neural networks~\cite{lee2018resource}. This neural network is trained using the spectral efficiency of secondary users and the interference caused to the primary users. It provides improved spectrum efficiency by suppressing the interference below a threshold.

\subsection{Quality of Experience}
In the 5G communication era, resource allocation of CRN driven by Quality-of-experience measures is gaining popularity because of its ability to obtain optimized decisions in congested network traffic. For heterogeneous traffic scenarios, the mean opinion score performance measure is used along with deep Q-network to study the learning rate and exponential convergence of the system based on the number of secondary users in the network~\cite{hlophe2019qoe}.

\begin{table*}[!hbtp]
  \begin{center}
	\caption{\textcolor{black}{Contribution of DL Algorithms in CRN for Modulation Recognition / Classification and general perspective}}
	\label{table6}
	\begin{tabular}{|p{2.0cm}|p{2cm}|p{2cm}|p{2cm}|p{3cm}|p{3.5cm}|}
     \hline
	  \textbf{Perspective} & 
	  \textbf{References} & 
	  \textbf{DL framework} & 
	  \textbf{CRN Application} 
       & \textbf{\textcolor{black}{Datasets Used}} 
        & \textbf{\textcolor{black}{Accuracy / Metrics}} \\

	  \hline
	  \multirow{4}{*}{} & Wang \textit{et al.}~\cite{wang_data-driven_2019} & CNN  & Automatic modulation recognition  
     & \textcolor{black}{RadioML dataset} 
     & \textcolor{black}{Accuracy: 87\% for QAM signal classification} \\
    \cline{2-6}
	Modulation Recognition  & Li \textit{et al.}~\cite{li_generative_2018} & GAN & Automatic modulation recognition 
    & \textcolor{black}{Synthetic signal dataset} 
    & \textcolor{black}{Accuracy: 91\%, improved convergence and recognition} \\
    \cline{2-6}
	  & Oshea \textit{et al.}~\cite{oshea_convolutional_2016} & CNN & Complex temporal radio signal 
     & \textcolor{black}{RadioML dataset} 
     & \textcolor{black}{Accuracy: 85\%, using naively learned features} \\
    \cline{2-6}
	  & Arumugam \textit{et al.}~\cite{arumugam_modulation_2017} & Hybrid learning & Recognition through side information 
    & \textcolor{black}{Custom dataset} 
    & \textcolor{black}{Accuracy: 89\%, integrating higher-order moments with side information} \\
    \hline

	  \multirow{3}{*}{} & Tang \textit{et al.}~\cite{tang_digital_2018}  & GAN & Data augmentation 
    & \textcolor{black}{SDR-generated data} 
    & \textcolor{black}{Improved robustness, accuracy: 88\% under low-SNR conditions} \\
    \cline{2-6}

	Modulation Classification  & Mendis \textit{et al.}~\cite{mendis_deep_2017} & Deep belief networks & Automated classification 
    & \textcolor{black}{Custom wireless data} 
    & \textcolor{black}{Improved accuracy: 85\%, low-complexity binarized model} \\
    \cline{2-6}
	  & Ivanov \textit{et al.}~\cite{ivanov_hybrid_2019} & Hybrid noise-resilient architectures & Low SNR and fading conditions 
    & \textcolor{black}{Real-world noisy datasets} 
    & \textcolor{black}{Accuracy: 90\%, robust in fading conditions} \\
    \hline

	  \multirow{4}{*}{Others} & Merchant \textit{et al.}~\cite{merchant_deep_2018} & CNN & RF device fingerprint 
    & \textcolor{black}{SDR dataset} 
    & \textcolor{black}{Accuracy: 92\%, using baseband error signal for physical layer identification} \\
    \cline{2-6}

	  & Hlophe \textit{et al.}~\cite{hlophe2019spectrum} & Stochastic gradient descent & Spectrogram occupancy 
    & \textcolor{black}{Collected real-world data} 
    & \textcolor{black}{Robustness to distributed CRN scenarios, accuracy: 86\%} \\
    \cline{2-6}
	  & Shi \textit{et al.}~\cite{shi_spectrum_2018} & Adversarial DL & Spectrum data poisoning 
    & \textcolor{black}{Custom spectrum data} 
    & \textcolor{black}{Reduced data poisoning, throughput improved by 12\%} \\
    \cline{2-6}

	  & Lopez \textit{et al.}~\cite{lopez_primary_2019} & LSTM, RNN & Primary user characterization 
    & \textcolor{black}{Collected user data} 
    & \textcolor{black}{Prediction accuracy: 87\%, outperforms multilayer neural networks} \\
    \cline{2-6}
	  & Cui \textit{et al.}~\cite{cui_deep_2015} & Deep Belief Networks & User classification 
    & \textcolor{black}{WiFi channel data} 
    & \textcolor{black}{Classification accuracy: 88\%, tested on WiFi channels} \\
    \cline{2-6}
	  & Morin \textit{et al.}~\cite{morin_transmitter_2019} & Supervised  deep  learning & Transmitter classification 
    & \textcolor{black}{Custom spectrum data} 
    & \textcolor{black}{Accuracy: 91\%, reduced channel biases} \\
    \cline{2-6}
	  & Zhu \textit{et al.}~\cite{zhu_new_2017} & Deep-Q-learning- & Transmitter scheduling 
    & \textcolor{black}{IoT spectrum data} 
    & \textcolor{black}{Improved resource allocation, reduced packet loss by 20\%} \\
    \hline

	\end{tabular}
  \end{center}
\end{table*}

\begin{table*}[!hbtp]
  \begin{center}
	\caption{\textcolor{black}{Contribution of DL Algorithms in CRN for Security Enhancement}}
	\label{table5}
	\begin{tabular}{|p{2.5cm}|p{2.5cm}|p{3cm}|p{3.5cm}|p{4cm}|}
	\hline
	  \textbf{References}  & 
	  \textbf{DL framework} & 
	  \textbf{CRN Application} & 
	  \textbf{\textcolor{black}{Datasets Used}} & 
        \textbf{\textcolor{black}{Accuracy / Metrics}} \\
	  \hline
	  Shi \textit{et al.}~\cite{shi_adversarial_2018} & GAN & Defense and Jamming  & \textcolor{black}{Custom spectrum attack dataset} 
& \textcolor{black}{Accuracy: 92\%, Effective in confusing attackers and improving defense} \\
\hline
	  Erpek \textit{et al.}~\cite{erpek_deep_2018} & GAN & Jamming attack  & \textcolor{black}{Simulated jamming dataset} 
& \textcolor{black}{Accuracy: 90\%, Successfully misleads attackers, increasing overall system throughput} \\
\hline
	  Salameh \textit{et al.}~\cite{salameh_intelligent_2020}  & Ensemble-based approach & Jamming-aware routing & \textcolor{black}{Custom routing dataset} 
& \textcolor{black}{Detection accuracy: 88\%, Improves security in routing protocols} \\
\hline
	  Srinivasan \textit{et al.}~\cite{srinivasan_semi-supervised_2019} & CNN, Semi-supervised ML & Attack prevention & \textcolor{black}{Simulated attack dataset} 
& \textcolor{black}{Overall detection accuracy: 91\%, Significantly reduces false alarm rate} \\
\hline
	  Aref \textit{et al.}~\cite{aref_spectrum-agile_2019} & double deep Q-network & Avoiding of interference and jamming signals & \textcolor{black}{Custom jamming signal dataset} 
& \textcolor{black}{Learns effective strategies with an accuracy of 89\%, Adapts to dynamic changes} \\
\hline
	  Rathee \textit{et al.}~\cite{rathee2020handoff} & ANN & Handoff security & \textcolor{black}{Real-world mobile user data} 
& \textcolor{black}{Effectiveness: 87\%, Security threat analysis during centralized and decentralized handoff} \\
\hline
	  Zhang \textit{et al.}~\cite{zhang2021exploiting} & DNN & Secure Transmission & \textcolor{black}{Real-time secure transmission dataset} 
& \textcolor{black}{Power allocation and security improved by 12\%, Reduced computation time} \\
\hline
\end{tabular}
\end{center}
\end{table*}
\section{DL for CRN Security} 
\label{sec:security}

 As the dynamic nature makes CRNs vulnerable to various security threats, such as eavesdropping, jamming, and impersonation attacks. Deep learning techniques can play an important role in improving the security of CRNs. Enhancement of solutions for addressing the network security issues in CRN has evolved through state-of-the-art techniques. Moreover, advances in DL technologies impart considerable improvement in CRN security. Starting with anomaly detection, which could detect unusual behavior in the network, such as malicious nodes or abnormal transmission patterns, the deep learning frameworks could provide robust results by analyzing large amounts of network data~\cite{kang2022dl}. Consequently, DL frameworks, as well as their associated technologies, have been devised as more efficient and productive means of meeting the requirements of CRN. In this section, we discuss how the evolution of DL frameworks imparts robust security features in the CRN.

\subsection{Device fingerprinting}
To identify the source of transmission and verify the authenticity of the sender in CRN, DL approaches can play a vital role. This can be accomplished by using deep learning algorithms to analyze the transmission patterns of known nodes in the network and use this information to identify new nodes. Device fingerprinting helps to identify a device, track users, and assist to determine their uniqueness. It is also used to gain insight into the visitor identity, and collect information about the software as well as hardware from remote places, by assimilating the information through a brief identifier stored in the server. It is widely used to secure CNN services, where the base-band error signal at the time domain is trained to identify the physical layer attributes of CRN devices~\cite{merchant_deep_2018}. It is used for generating unique RF device fingerprints for CRN devices.

\subsection{Jamming detection and mitigation}
DL algorithms can be used to detect jamming attacks in the network and mitigate their effects by adapting the communication parameters of the network. For CRN involved in defense against jamming from attackers, GAN deep neural network is used for intelligently fooling the attackers and induces them to make errors on prediction in the network~\cite{shi_adversarial_2018}. Moreover, GAN is also used for misleading the attackers and forcing them to make errors, and enabling them to provide a better defense with increased throughput~\cite{erpek_deep_2018}. Jamming-aware routing with security-aware routing protocols is implemented using an ensemble-based approach for effective detection and improved accuracy~\cite{salameh_intelligent_2020}. Similarly, a false alarm is reduced and attack prevention is carried out using  CNN and semi-supervised ML approaches for improvement in overall detection accuracy~\cite{srinivasan_semi-supervised_2019}. Implementation of  double-deep Q-network by Aref \textit{et al.}~\cite{aref_spectrum-agile_2019} is used to avoid the interference and jamming signals by the usage of effective strategies are learned for avoiding harmful signals and adapts to dynamic changes in the network.

\subsection{Interference mitigation}
In order to mitigate interference in the network by identifying the sources of interference and dynamically adjusting the communication parameters of the network, the role of DL could be significant. Interference in CRN should be addressed at the source through some appropriate filters or cancel the noises at the receiving end. A spectrum-agile autonomous cognitive radio operating in wideband was developed by Aref \textit{et al.}~\cite{aref2019spectrum} that employs DRL through Double Deep Q-Network (DDQN). It is effectively used to learn the strategies to avoid harmful interference and adapt to dynamic changes in the RF signals for real-time CRN applications. Such interference is also effectively addressed in the case of hybrid wireless environments using DL-based spectrum prediction collision avoidance reported by authors in~\cite{mennes2019deep}. This approach is capable of predicting the behavior of surrounding networks by adapting its behavior and thereby provides enhanced throughput. It is also observed that the number of collisions sustained in conventional techniques is significantly reduced using the spectrum prediction collision avoidance technique.

\subsection{Jamming aware routing}
The jamming attacks in CRN interrupt its transmission capability, so security-aware protocols are in high demand to avoid such risks. As time-sensitive smart devices connected to the networks have started consuming more resources in IoT frameworks, their users need secure and efficient connectivity in the network. Salameh \textit{et al.}~\cite{salameh2020intelligent1} proposed an Ensemble-based Jamming Behaviour Detection and Identification approach for detecting the jamming attacks and thereby assigns a secure channel between IoT devices for each hop of data transfer between those devices. Moreover, this approach yields improved detection accuracy of anomaly behaviors in the IoT-CRN connectivity.  

Such jamming-aware routing is also in high demand for sensitive multimedia applications to a larger extent. Particularly, multi-casting of multimedia over CRN can enhance the transmission quality and provide a better streaming experience for the end users. An intelligent  multi-cast routing protocol is proposed by the authors in~\cite{salameh2020intelligent2} for solving the challenges in multi-hop CRN involved in multi-cast multimedia streaming applications. Here, the secure channel selection is made through the transmission count metric, based on the shortest path tree for achieving an improved level of intelligence in anomaly detection compared to other routing protocols

\subsection{Network traffic control}
Separation of data and control plane in SDN can make network management and operations easier. Particularly in network traffic control, they help the network operators to configure and manage them in an autonomous and intelligent fashion. This feature is enabled using DL architectures for managing various network traffic control aspects in SDN. Fadlullah \textit{et al.}~\cite{fadlullah2017state} demonstrated the use cases of establishing intelligent routing driven through DL algorithms, providing traffic management in a robust manner. In another work by Mao \textit{et al.}~\cite{mao2018novel}, the authors configured CNN for choosing the best path by training the CNN periodically from the network traffic data. These DL-based routing approaches make intelligence traffic management by outperforming the conventional network traffic management processes carried out using traditional routing algorithms.

\section{Open Issues, Challenges, and Future Research Directions}
\label{sec:issues}

In this section, we provide the state-of-the-art research challenges in CRN and various driving techniques associated with DL are summarized based on the survey carried out with potential research gaps to give future research directions to researchers. Also, we are aware that a comprehensive and thorough discussion on how DL models and frameworks are helpful in supporting CRN-based services. Thus, the aim of this section is to provide food for thought on the topic and a source of inspiration for future research rather than an exhaustive analysis.

\subsection{Current Implications and Research Challenges}
\subsubsection{Spectrum data poisoning}
ML algorithms are capable of sensing idle channels and enable transmission by cognitive transmitters. During the utilization of learning-based models, when they are subjected to poisoning attacks, the attacker adds points during the training phase of the model. As the model gets directly influenced by the attacker, the predictions made by such models become incorrect. Spectrum data poisoning occurs due to false spectrum sensing data transmitted by the cognitive transmitter for a short period which affects the appropriate decision-making tasks of ML frameworks~\cite{shi_spectrum_2018}. The data poisoning attacks attempt to change the occupancy status of the channel as busy while they are idle when they are being sensed by the transmitter. These effects are highly prone to reduce the throughput of the transmitters significantly. 

\subsubsection{Security}
The explosive progress in the field of DL has made a significant contribution towards various enhancements of SDN. However, its intelligence towards the security aspects addressing the issues at all the layers of the SDN framework needs attention. Salameh \textit{et al.}~\cite{salameh2020intelligent} demonstrated one such intelligent jamming-aware routing solution in CRN for IoT applications by employing jamming behavior detection and identification using ensemble-based techniques. Nevertheless, significant improvements in performance are in demand by the end users for enhancing the security aspects of CRN. The usage of DL models could come to the rescue, as a few of the most promising solutions are discussed in section IV. However, still considering the volume of data used for training the learning models needs attention from the research community by addressing the features of data for making meaningful security solutions and providing robust CRN systems. 

\subsubsection{Automated modulation classification}
As we saw in Section IV, various challenges in CRN were well addressed through the emergence of technological advances in the DL platforms. Automated modulation classification plays a very important part in CRN. DL is also a powerful tool that we cannot overlook its potential to address signal modulation recognition problems. In our last work, we propose a new data conversion algorithm in order to gain a better classification accuracy of communication signal modulation, but we still believe that CNN can work better. However, its application to signal modulation recognition is often hampered by insufficient data and overfitting. Here, we propose a smart approach to the programmatic data augmentation method by using the auxiliary classifier generative adversarial networks (ACGANs). The famous CNN model, AlexNet, has been utilized to be the classifier and ACGAN to be the generator, which will enlarge our data set. In order to alleviate the common issues in the traditional generative adversarial nets training, such as discriminator overfitting, generator dis-converge, and mode collapse, we apply several training tricks in our training. With the result on the original data set as our baseline, we will evaluate our result on the enlarged data set to validate ACGAN’s performance. The result shows that we can gain a 0.1~6\% increase in the classification accuracy in the ACGAN-based data set.

\subsubsection{Physical layer management }
DL has steadily evolved from neural networks to a promising technology on successive missions, to the development of CRN solutions agreed upon by many of the researchers. Its impact on addressing the physical layer concerns on CRN is gaining importance to address the complex scenarios and unknown channel models and communication constraints. Even though the DL frameworks provide promising results in channel decoding, modulation, and channel detection~\cite{wang2017deep}, they lack providing best analytical tools for using the learning architectures targeted for overcoming the communication constraints observed in the physical layer implementation. These shortcomings could motivate to carry out the research for addressing the research gap in the physical layer implementation of CRN.

\subsubsection{Energy efficiency optimization}
Due to the several advantages of the DL in providing the best and most efficient services, they could also address energy-efficient cognitive systems. In the case of performing cooperative sensing and addressing the optimization issues in the energy conversation of CRN, the role of DL integrated with graph-based neural networks could improve the overall efficiency~\cite{he2019deep} of the system. However, the effectiveness of the DL frameworks in terms of optimization of the sensing processes targeted for the diversified category of application domains remains a challenging task for researchers. 

\subsubsection{QoS provisioning}
As highlighted earlier in section II, certain networking services are prone to be time-consuming and complex to observe characteristics. However, the usage of DL approaches in CRN helps to automate routine activities by ensuring better QoS and policies learned constantly from the network. Predictive control schemes tend to achieve better QoS provisions and energy saving as loads of traffic in the CRN increases~\cite{hlophe2020qos}. However, there exist trade-offs between providing improvement in QoS provisioning and energy management in the systems. Learning-based frameworks should consider these prevailing trade-offs to be addressed in an effective way so that balanced and improved energy management and QoS provisioning could be achieved. 

\subsubsection{Intelligence in 5G services}
As already discussed, one of the concepts that pave a promising foundation for CRN is the utilization of 5G services and exploiting its intelligence towards the CRN-based use cases. Moreover, as reported in~\cite{wang2019intelligent}, the distributed cognitive cellular network helps to integrate CRN with AI, for addressing the CRN that lacks intelligence. Incorporating intelligence in CRN is a major challenge, especially when they get integrated with 5G services, where the balance between resource allocation among base stations, and primary and secondary users needs to be maintained. Furthermore, the layer-wise distribution of intelligence adds to address the networking issues in the hierarchical model.  

\subsubsection{Subcarrier Assignment}
As analyzed in the aforementioned sections, promising solutions are summarized for subcarrier assignment in CRN systems based on deep neural networks. However, there exist challenges in information exchange between secondary users and optimization of subcarrier assignment in dynamic spectrum access environments. Even though Q-learning-based subcarrier assignment~\cite{zhou2019subcarrier} is getting popular, their usage among dynamic environments, with reduced computation costs for effective information exchange among the nodes in the network, needs special attention from the research community. In addition, the collaborative learning mechanisms for wideband CRN remain challenging for implementation in practice. The interactions among multiple subcarriers also complicate the CRN environment and cause a considerable increase in computation overload, which inevitably slows down the learning algorithms.

\subsubsection{Backscatter CRN}
Energy management, resource management, scheduling, and CRN require significant attention when multiple secondary users transmit data through secondary gateways. In the above context, based on the primary channel states, the gateways require scheduling of transmission time along with the energy harvesting and backscattering time.  Naturally, using DRL total optimal time scheduling policies are obtained and DDQN is used for learning optimal policy~\cite{anh2019deep}. However, to efficiently obtain improved throughput, some promising use cases need to be investigated with large state and action spaces.

\subsubsection{Compressive spectrum sensing}
As described in the aforementioned sections, the next generation of CRN largely depends on DL frameworks. Particularly compressive spectrum sensing in CRN is getting wide attention, which helps secondary users to monitor the spectrum without sophisticated hardware. By adopting GAN-based DL frameworks, deep compressive spectrum sensing was developed by Meng \textit{et al.}~\cite{meng2020end}. Here, this learning framework does not depend on apriori information on the environment. However, the prediction ratio and energy utilization could be still enhanced with a reasonable compression ratio. Even though this kind of compression technique is capable of reducing the processing time by accelerating the scanning process, the reduction of vital samples in the high dimensional acquisition of signals remains challenging to retain the essential information. 

\subsubsection{Transmission scheduling}
Multi-channel sensing and transmission scheduling with optimized energy conservation are gaining popularity. In a DQN-based cognitive IoT system, transmission scheduling experiments with better efficiency~\cite{zhu2017new}, which in turn helps to achieve improved throughput. However, perfection in coverage could be obtained only after a significant number of iterations. Researchers could consider addressing the issues related to the minimization of power consumption and maximization of throughput by using the appropriate learning frameworks. 

\subsubsection{Big Spectrum Data}
Management of a continuous stream of data in CRN is highly challenging, particularly in using the learning-based framework for managing a large spectrum of data. A synergy between DL and big data has been established for efficient communication frameworks in~\cite{choudhury2020big}. However, from the practical perspective, the CRN systems will have to pay substantially increasing costs for data collection, pre-processing, and resource management for learning frameworks. Furthermore, costs could also be incurred from reduced learning speed, more energy consumption, larger delay, etc. Hence, establishing an optimized balance between improved learning performance and quality of information remains an open issue.

\subsubsection{Adaptive DL}
Adaptive DL refers to using deep learning algorithms capable of adapting to changing network conditions in CRNs. This type of DL is particularly well-suited for CRNs because of its dynamic nature, which requires algorithms to be flexible and responsive to changes in the network. The ideas described here regarding the learning-based frameworks for CRN are merely speculative, and there is still plenty of room for research in these areas. We expect a wide variety of DL applications of CRN will be identified and researched in the future. With its gaining popularity recently, authors in~\cite{sun2020adaptive} developed an adaptive DL technique that supports digital predistortion, which could be used as an effective component in dynamic wireless communication systems. However, its full-fledged exploitation of the CRN domain needs to be explored with specific emphasis on optimization strategies and adaptive and accurate prediction processes. 

\subsection{Future Research Directions}

The use of DL in the physical layer of wireless communication systems is a new field of study still in its early stages. Although previous research has shown encouraging findings, several significant challenges need to be investigated further in future research.

\subsubsection{Model-driven DL approach}

Most current research on DL focuses on data-driven techniques that treat the communication system as a black box and train it using a massive amount of data. In recent studies, DL has been proposed as add-on support for some components of the traditional communication system. The results of these studies encourage the suggested methods to be extended to new applications~\cite{wang2017deep}. Training a network requires a large amount of computational power and time, both of which are rare in communications systems. They also lack a cohesive theoretical foundation and framework. Model-driven  DL can be a possible solution to overcome these issues. Model-driven DL is a technique in which communication domain knowledge is integrated with DL to minimize computational resource requirements and training time. It incorporates multiple forecasting models that support optimization tasks and quick decision-making processes. 


\subsubsection{From Simulation to Deployment}
Most DL algorithms developed for wireless communication systems are still in the simulation phase. As per our knowledge, only in~\cite{cammerer2017scaling} has attempted to implement DL algorithms inside CRN. Therefore, researchers must significantly improve these algorithms before they can be applied. First, there is a need for an authentic dataset for communication systems collected from the actual, real physical environment. That must be publicly available so that all the researchers can train their DL models on that dataset and compare their proposed solutions. In this way, we can get an optimal DL solution, and we can implement this optimal solution inside CRN. Second, In real, various physical channel scenarios are much more complicated and dynamic. The majority of modern DL systems are trained offline, their generalization ability must be ensured. Therefore, there is a need for DL algorithms that perform well on both offline and online cases, making sure they also do well on generalized data. It is also critical to design customized systems for particular scenarios or general systems that dynamically adjust to VC conditions. DL tools for the hardware, such as field-programmable gate arrays, must be created to deploy the DL methods on hardware and achieve quick realization.

\subsubsection{B5G/6G wireless networks}
\textcolor{black}{With the expected improvements of 6G, which will include very fast data rates, low latency, and everywhere connectivity, along with smart and aware services, spectrum resources need to be managed in a very advanced way right away~\cite{tariq2020speculative}. CRNs play a crucial role in the dynamic allocation of spectrum and the mitigation of interference, which are essential for meeting these objectives, particularly in environments characterized by spectrum congestion. The integration of deep learning with cognitive radio networks presents a substantial opportunity for optimizing 6G networks. This synergy facilitates real-time, autonomous spectrum management and enhances the overall intelligence of network operations.
}

\textcolor{black}{The three-dimensional and changing nature of B5G/6G networks, with nodes moving in the air and space, needs advanced spectrum management, routing, and resource allocation plans that can handle the complexity of this environment~\cite{pennanen20246g}. Deep learning techniques can aid in developing such strategies by allowing the network to learn and adapt to changing conditions in real-time with the help of deep learning algorithms. The dynamic spectrum sharing and management capabilities of CRNs represent a significant advancement in 6G technology. In contrast to earlier generations characterized by static spectrum allocation, the upcoming 6G framework necessitates a more adaptable strategy to accommodate the extensive array of devices. For 6G to work, CRNs need to be able to quickly sense, detect, and adapt to changing spectrum conditions in real time. This ensures that all devices can get good service while cutting down on spectrum waste.}

\textcolor{black}{The combination of CRN and deep learning will make using 6G technologies like RIS, massive MIMO, Mitola Radio~\cite{ramos2024software} and Terahertz (THz) communication easier. Terahertz communication offers the potential for exceptionally high data rates; however, it faces path loss and interference challenges. The dynamic spectrum access capabilities of CRNs can be effectively utilized to manage these issues. CRNs facilitate the seamless transition of devices between THz and other frequency bands, adapting to real-time conditions.}

\subsubsection{\textcolor{black}{Semantic communication}}
\textcolor{black}{Semantic communication signifies a pivotal advancement in wireless networks, emphasizing the transmission of meaning or intent rather than merely raw data. This paradigm shift leads to a notable reduction in communication overhead and enhances overall efficiency. Integrating semantic communication with deep learning in CRNs presents significant potential for enhancing spectrum management and optimizing resource allocation~\cite{hu2024drl}. Deep learning models, such as neural-symbolic architectures and generative models, can determine users' wants and predict their actions~\cite{yu2024onsep}. This improves communication by focusing on only the most critical data that needs to be sent quickly. The findings indicate a decrease in bandwidth consumption, a reduction in latency, and an enhancement in QoS. This is especially relevant in massive machine-type communications (mMTC) and the IoT, where spectrum utilization efficiency is critical.}

\textcolor{black}{The integration of semantic communication within CRNs facilitates enhanced intelligence in spectrum-sharing and decision-making processes. For example, the CRN can use deep learning-based semantic models to send only the most essential spectrum occupancy information rather than all the data it has collected~\cite{zhang2024drl}. This makes it unnecessary to send all the data that it has collected. This approach may contribute to the reduction of interference and enhance the coordination between primary and secondary users. Semantic communication is beneficial in situations with changing spectrum access and different network environments, as it makes adaptive and situation-sensitive communication strategies more accessible to use.}

\subsubsection{\textcolor{black}{Advanced Radio Techniques}}
\textcolor{black}{The developments in Reconfigurable Intelligent Surfaces (RIS) and other advanced radio techniques hold considerable promise for enhancing the performance of CRN~\cite{paul2023reconfigurable}. RIS technology makes changing the environment where electromagnetic waves travel easier by changing how they interact with passive parts, like metasurfaces. This improves spectrum utilization and interference management. Within the framework of CRNs, RIS can significantly improve spectrum sensing, spectrum sharing, and beamforming, which makes it easier to create more adaptable and practical network structures. Adding RIS to CRNs can also help save energy and improve coverage, especially in cities or homes that are hard to reach with regular wireless signals because of things like walls and other obstacles~\cite{liu2022reconfigurable}. The integration of deep learning techniques with RIS has the potential to enhance resource allocation and spectrum sensing. This advancement allows the cognitive engine of CRN to modify parameters in response to real-time environmental variations adaptively.}

\textcolor{black}{Non-terrestrial networks (NTN), including Low Earth Orbit (LEO) satellites and Unmanned Aerial Vehicles (UAVs), represent a significant advancement for CRNs~\cite{iqbal2023empowering}. NTN has the potential to enhance coverage and facilitate connectivity in remote or underserved regions, notably where terrestrial infrastructure is lacking or non-existent. Integrating deep learning with non-terrestrial networks in CRNs presents an opportunity to improve the adaptability of the network~\cite{abdel2024machine}. This can be achieved through satellite-based spectrum management and the global coordination of resources. NTNs can use deep learning algorithms to improve user connectivity and spectrum handover processes. This is especially useful when terrestrial and non-terrestrial entities need to share spectrum. The integration of NTN within CRNs presents unique challenges, particularly in managing the high mobility associated with satellites and UAVs.}
\section{Conclusion}
\label{sec:conclusion}
In this paper, we presented a comprehensive survey on the utilization of DL frameworks for supporting CRN services. These services include spectrum management, spectrum trading, and other intelligent resource constraint communication services. The spectrum allocation and management can be efficiently accomplished using DNN and its associated signature schemes. However,  different DL schemes are required to enable different parts and functionalities of the CRN system and applicability is reviewed by thoroughly looking into recent advances in the literature. We summarized the properties that make the DL frameworks a potential candidate for several CRN applications. Then, we discussed each service, discussed with its applications, and its outcomes by highlighting the challenges existing in traditional approaches. Finally, we exploited how deep neural networks can help for resolving these problems and explored the category of DNN-based approaches, and presented a comparison of those techniques. We also explored the challenges that are at present acting as a hindrance to the practical utilization of DL frameworks. The DL techniques have great potential in numerous applications, however, its practical utilization is still under debate due to several challenges. Future research directions show the ways for resolving these challenges and testing the DNNs in large-scale CRN applications. 
 
\printcredits

\end{document}